\documentclass[aps,twocolumn,superscriptaddress]{revtex4}
\usepackage{graphicx}

\newcommand{\ket}[1] {\left| #1 \right\rangle}

\begin{document}

\title{Nonlinear Landau-Zener-St\"uckelberg-Majorana problem}

\author{Sahel Ashhab}
\affiliation{Advanced ICT Research Institute, National Institute of Information and Communications Technology (NICT), 4-2-1, Nukui-Kitamachi, Koganei, Tokyo 184-8795, Japan}
\author{Olga A.~Ilinskaya}
\affiliation{B.~Verkin Institute for Low Temperature Physics and Engineering, Kharkiv 61103, Ukraine}
\author{Sergey N.~Shevchenko}
\affiliation{B.~Verkin Institute for Low Temperature Physics and Engineering, Kharkiv 61103, Ukraine}
\affiliation{School of Physics and Technology, V.~N.~Karazin Kharkiv National University, Kharkiv 61022, Ukraine}

\date{\today}

\begin{abstract}
In the standard Landau-Zener-St\"uckelberg-Majorana (LZSM) problem, the bias sweep rate and gap are both time independent and fully characterize the LZSM problem. We consider the nonlinear LZSM problem, in which at least one of the two characteristic parameters varies as the system traverses the avoided crossing region. This situation results in what could be thought of as a more accurate description of any realistic situation as compared to the idealized linear LZSM problem. We consider both the case of perturbative nonlinearities, where the nonlinearity adds small corrections to the linear problem, and the case of essential nonlinearities, where the sweep and/or minimum-gap functions are qualitatively different from those of the linear LZSM problem. In the case of perturbative nonlinearities, we derive analytic expressions for the LZSM transition probability based on the Dykhne-Davis-Pechukas (DDP) formula, taking into account the leading corrections to the standard LZSM formula. We compare the derived approximate expressions with numerical simulation results and comment on the validity of the approximations. In particular, if the nonlinear term is small in comparison to the linear term throughout the finite duration of the avoided crossing traversal, the perturbative approximation is valid. Our results also provide information about the validity of the DDP formula. In addition to reviewing cases of essential nonlinearity treated previously in the literature, we analyze the case of an essentially nonlinear sweep function that describes an almost square pulse.
\end{abstract}

\maketitle

\section{Introduction}
\label{Sec:Introduction}

The Landau-Zener-St\"uckelberg-Majorana (LZSM) problem \cite{Landau,Zener,Stueckelberg,Majorana} (see also the review \cite{ShevchenkoReview}) deals with the quantum dynamics of two quantum states when some system parameter is varied such that the two corresponding energy levels approach each other, experience an avoided crossing and move apart as the variable parameter continues its variation. Despite its simplicity, the LZSM problem applies to a remarkably broad set of physical phenomena ranging from atomic collisions and chemical reactions to the operation of quantum computing machines. The dynamics of LZSM transitions can also be used for the control of quantum systems \cite{Ivakhnenko,ShevchenkoBook,Nakamura}. Parameter variations can also lead to other effects in two-level systems, such as motional narrowing and averaging \cite{SilveriReview}.

The probability to make a transition between the two quantum states in the LZSM problem was derived independently by Landau, Zener, St\"uckelberg and Majorana \cite{Landau,Zener,Stueckelberg,Majorana} (see also \cite{DiGiacomo}). In its basic formulation, a system parameter is varied linearly from an infinitely large negative time to an infinitely large positive time. This idealized assumption renders the problem exactly solvable, resulting in the well-known LZSM formula.

If the problem deviates from the idealized scenario of a linear sweep function and fixed gap, the problem is in general not exactly solvable. There have been several studies over the years on variants of the LZSM problems. In particular, a number of special cases with specific sweep functions have been shown to allow exact solutions. These exactly solvable special cases, however, were generally obtained by identifying functions that satisfy certain mathematical conditions and hence allow analytical treatment. As a result, they do not necessarily represent better approximations for realistic physical systems in comparison with the original, linear LZSM problem.

In this work we calculate what can be considered the leading corrections to the LZSM formula arising from a weak nonlinearity in the sweep function or a weak time dependence of the gap. Quantum technologies have made remarkable advances in recent years. Higher-order corrections to the idealized LZSM transition probability, which might have been of mostly academic interest in the past, could now be measured and possibly utilized in future practical applications. This situation makes it imperative to have more accurate approximations for the LZSM probability in a nonlinear setting. In addition to our analysis of the weak nonlinearity case, we consider the power and error functions as sweep functions in the limiting case when they are almost square pulses.

The paper is organized as follows: In Sec.~\ref{Sec:Problem}, we introduce the idealized LZSM problem and how a nonlinearity is added to the problem. In Sec.~\ref{Sec:DDP}, we describe the Dykhne-Davis-Pechukas (DDP) formula, which can be applied to a general LZSM problem. In Sec.~\ref{Sec:PreviousWork}, we review previous work in the literature on the nonlinear LZSM problem. In Sec.~\ref{Sec:PerturbativeNonlinearity}, we treat the case of a weak nonlinearity: we derive analytic expressions for the corrections to the LZSM formula when we include the leading-order nonlinear correction term to the sweep function, and we test the limits of the perturbative formulae with numerical calculations. In Sec.~\ref{Sec:OtherCases}, we consider a few additional interesting cases of the nonlinear LZSM problem. Section \ref{Sec:EssentialNonlinearity} gives results for an essentially nonlinear sweep function that describes an almost square pulse. In Sec.~\ref{Sec:VariableDelta}, we consider the case of a time-dependent gap and show that a general problem with a varying gap can be transformed to one with a fixed gap but modified sweep function. Section \ref{Sec:Conclusion} contains concluding remarks. In Appendix~A, we give a detailed calculation of the LZSM probability by the DDP method for a sweep function with a weak quadratic  nonlinearity. Appendix~B is devoted to the calculation of the LZSM probability for a sinh function. Appendix~C contains the detailed calculation for eliminating the time dependence of the gap, in addition to a perturbative formula for the case of a time-dependent gap.

\section{Linear and nonlinear LZSM problems}
\label{Sec:Problem}

The linear LZSM problem pertains to the dynamical evolution of a two-level quantum system under a specific type of temporal variation in the Hamiltonian. The dynamics of the two-amplitude state vector $\ket{\psi}$ is governed by the Schr\"odinger equation 
\begin{equation}
i\frac{d\ket{\psi}}{dt} = H \ket{\psi}
\label{Eq:TDSE}
\end{equation}
($\hbar=1$), where the Hamiltonian is given by
\begin{equation}
H = \frac{1}{2} \left( \begin{array}{cc} \epsilon(t) & \Delta \\ \Delta & - \epsilon(t) \end{array} \right),
\label{Eq:LinearLZHamiltonian}
\end{equation}
and the sweep function
\begin{equation} \label{LinearSweepFunction}
\epsilon(t)=v t, 
\end{equation}
with $v$ and $\Delta$ being the characteristic parameters of the LZSM problem. Physically $v$ is the sweep rate of the two diabatic state energies relative to each other, and $\Delta$ is the gap (i.e.~minimum separation) between the energies of the adiabatic states. Both parameters can be taken positive without any loss of generality. We shall refer to the $t$ value at which the adiabatic energy levels are closest to each other as the crossing point.

It should be noted that any linear Hamiltonian, i.e.~$H(t)=A\times t + B$ with Hermitian $2 \times 2$ matrices $A$ and $B$, can be transformed into the form given in Eq.~(\ref{Eq:LinearLZHamiltonian}), provided that $A$ is not proportional to the identity matrix. Working in a basis that diagonalizes $A$ makes the off-diagonal matrix elements time independent. Any imaginary part in the off-diagonal matrix elements can then be eliminated by multiplying the basis states with appropriate phase factors. The crossing point between the diagonal matrix elements can be set to $t=0$ via a shift in the time variable. Any non-antisymmetric component in the diagonal matrix elements can then be ignored, since an overall energy shift does not affect the LZSM dynamics.

The main quantity that is evaluated in the LZSM problem is the transition probability between the two quantum states as a result of traversing the avoided crossing region (i.e.~the region around $t=0$). It is therefore usually assumed that at the initial time $t\rightarrow -\infty$ the quantum state is given by $(\psi_{\uparrow}, \psi_{\downarrow}) = (1,0)$, or alternatively (0,1), and the goal is to determine the probabilities $|\psi_{\uparrow}|^2$ and $|\psi_{\downarrow}|^2$ at the final time $t\rightarrow \infty$, although some studies have considered the case of a quantum superposition initial state \cite{Kofman}. These quantities represent the probabilities that the quantum system will stay in its initial state or make a so-called nonadiabatic LZSM transition. The probability $P_{\rm LZSM}$ to remain in the same diabatic state, i.e.~$(\psi_{\uparrow}, \psi_{\downarrow}) = (1,0)$, is given by the well-known formula (see e.g.~Ref.~\cite{ShevchenkoReview}):
\begin{equation}
P_{\rm LZSM} = |\psi_{\uparrow}(t\rightarrow\infty)|^2 = e^{-2\pi\delta},
\label{Eq:P_LZ}
\end{equation}
where 
\begin{equation}\label{Eq:AdiabaticityParameter}
	\delta=\frac{\Delta^2}{4 v}
\end{equation}	
is the adiabaticity parameter.

\begin{figure}[h]
\includegraphics[width=5.5cm]{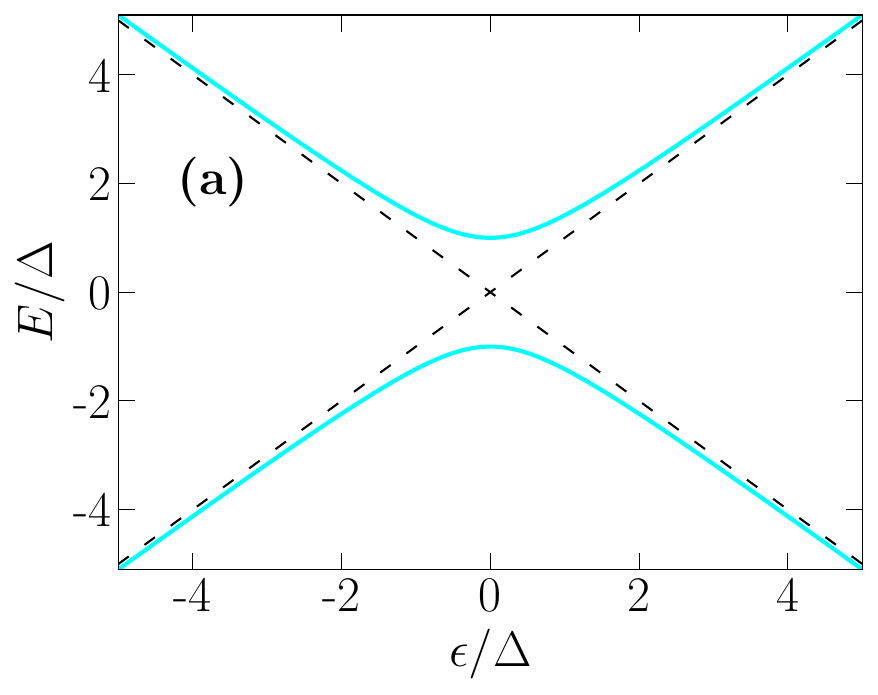}
\includegraphics[width=5.5cm]{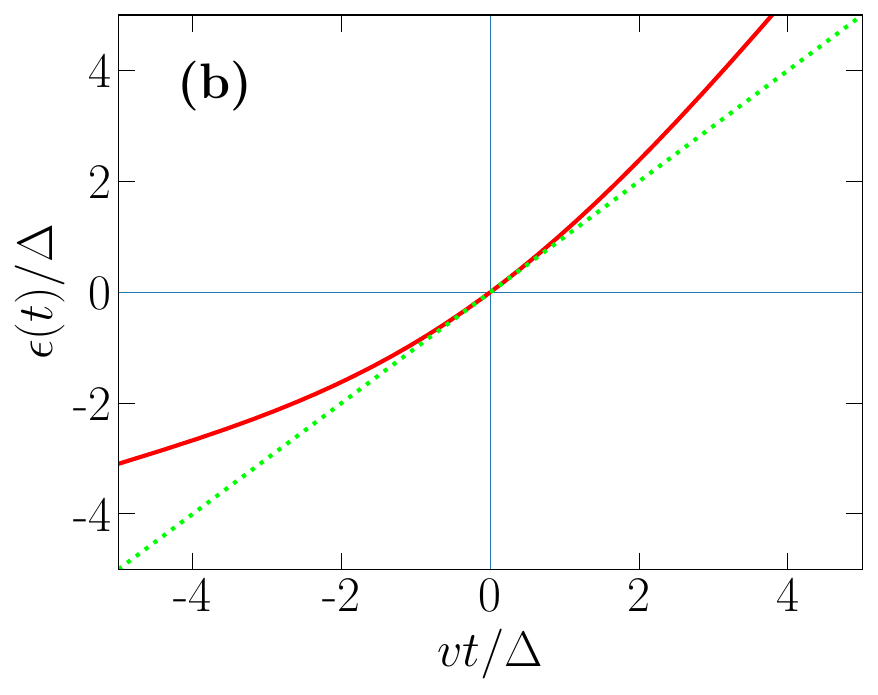}
\includegraphics[width=5.5cm]{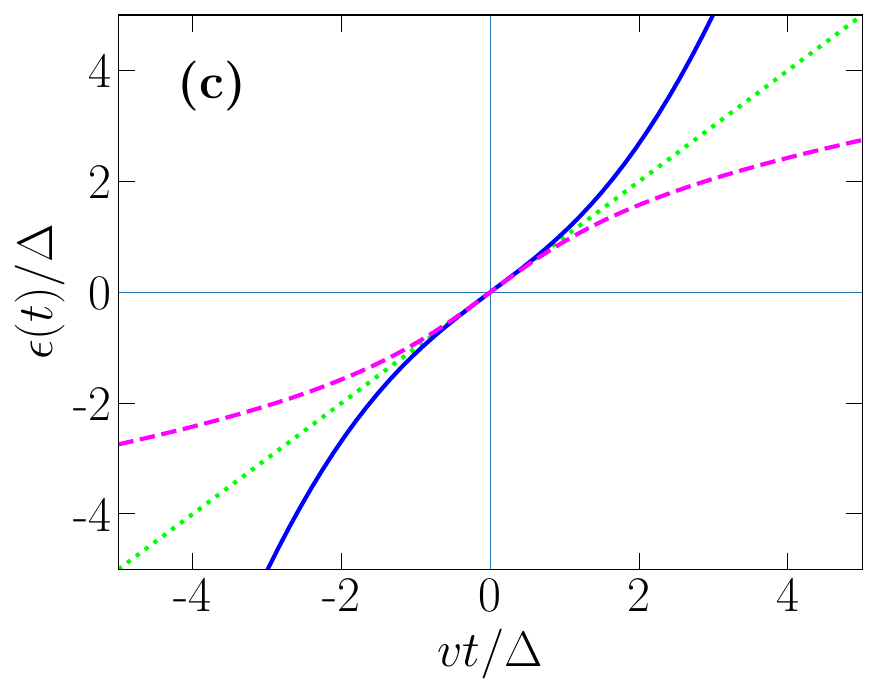}
\caption{Schematic diagrams of the energy levels and sweep functions of nonlinear LZSM problems. Panel (a) shows the energy levels as functions of the bias $\epsilon$, which is generally a function of time. As in the linear LZSM problem, the instantaneous energy levels (solid cyan lines) exhibit an avoided crossing. The dashed black lines show the energies of the diabatic states, which asymptotically approach the instantaneous energy eigenstates at $\epsilon\rightarrow\pm\infty$. In Panel (b), the solid red line corresponds to a nonlinear sweep function that contains both a linear term and a quadratic term: specifically, $\epsilon(t) = v t \times [1+0.5\tanh (vt/10 \Delta)]$. Panel (c) shows two examples of time-antisymmetric sweep functions: the solid blue line corresponds to the superlinear function $\epsilon(t) = v t \times [1+0.2(vt/\Delta)^2]^{1/2}$, while the dashed magenta line corresponds to the sublinear function $\epsilon(t) = v t \times [1+0.4(vt/\Delta)^2]^{-1/4}$. For comparison, the dotted green lines in Panels (b) and (c) show the linear sweep function $\epsilon(t)=vt$. The gap $\Delta$ is used as an energy unit for all the axes in this figure. }
\label{Fig:Schematic}
\end{figure}

The LZSM problem described above can, in some sense, be considered the simplest model for a quantum system undergoing nonadiabatic transitions by traversing an avoided crossing region. In particular, the LZSM problem has the appearance of an idealized linear model in which the parameters $v$ and $\Delta$ remain constant from $t\rightarrow -\infty$ to $t\rightarrow \infty$. In reality one would expect these parameters to vary over time for any actual physical setup, with $v$ being defined as the instantaneous time derivative of $\epsilon$, i.e.
\begin{equation}\label{v-on-t}
v(t)=\dot{\epsilon}(t).
\end{equation}
Note that throughout the manuscript a time-dependent $v$ will be defined by Eq.~(\ref{v-on-t}), and not by Eq.~(\ref{LinearSweepFunction}). On the other hand, approximating $v$ and $\Delta$ by constants for an infinite amount of time is not as bad as it might seem at first sight, because the LZSM transition dynamics takes place only during the traversal of the avoided crossing, as can be seen from a plot of the state probabilities as functions of time (see e.g.~Fig.~3 in Ref.~\cite{ShevchenkoReview}). The corresponding duration of the crossing process can, as a rough estimate, be defined as
\begin{equation}
\tau_{\rm LZSM} = \frac{1}{\sqrt{v}} {\rm max} \left( 1, \frac{\Delta}{2\sqrt{v}} \right).
\label{Eq:CrossingTime}
\end{equation}
The values of $v$ and $\Delta$ long before and long after the crossing point are almost irrelevant. For this reason, using only the values that $v$ and $\Delta$ take at the avoided crossing point gives a good approximation for the LZSM transition probability in realistic problems. Nevertheless, it is natural to expect that $v$ and $\Delta$ will generally vary in time even during the traversal period, and any such temporal variations of $v$ and/or $\Delta$ can be expected to affect the LZSM probability.

A simple picture of how the nonlinearity enters the problem is given by the case where an external parameter, e.g.~an externally applied field $E$, is varied linearly in time ($E=vt$) but the bias function $\epsilon$ is a nonlinear function $f(E)$ of the field. The bias function can then be expressed as
\begin{equation}
\epsilon(t) = f(vt) = vt + \frac{\chi_2}{2! \Delta} \left( vt \right)^2 + \frac{\chi_3}{3! \Delta^2} \left( vt \right)^3 + \cdots.
\label{Eq:NonlinearEpsilonExpansion}
\end{equation}
Here $\chi_2$ and $\chi_3$ are the quadratic and cubic nonlinearity parameters, respectively. The factors of $\Delta$ in the denominators are used to make the coefficients $\chi_n$ dimensionless. Corrections from nonlinear terms in physical problems typically decrease in importance as we go to increasingly high orders. In other words, one would intuitively expect that, for a problem with a weak nonlinearity, the quadratic term will produce the main correction to the transition probability, followed by the cubic term and so on. We shall show below that the quadratic nonlinearity leads to an especially small correction, which can make the cubic nonlinearity important even in the presence of a quadratic nonlinearity.

Figure \ref{Fig:Schematic} shows schematic diagrams of the energy level structure and various nonlinear sweep functions. The nonlinear function plotted in Fig.~\ref{Fig:Schematic}(b) contains a quadratic term, which is the leading-order correction to the linear approximation:
\begin{equation}
\epsilon(t) = v t \times \left(1+\alpha\tanh{\frac{v t}{\tilde{\alpha} \Delta}}\right) \approx 
v t+\frac{\alpha}{\tilde{\alpha}}\frac{(v t)^2}{\Delta}.
\end{equation} 
Figure \ref{Fig:Schematic}(c) shows examples of sweep functions that do not contain a quadratic term: a superlinear function,
\begin{eqnarray}\label{superlinear}
	\epsilon (t) &=& vt \left( 1 + \lambda t^{2} \right)^{1/2}\\
	&\approx& v t \left(1+\frac{\lambda t^2}{2}\right),\nonumber
\end{eqnarray}
and a sublinear function,
\begin{eqnarray}\label{sublinear}
 	\epsilon (t) &=& vt \left( 1 + 2 \lambda t^{2} \right)^{-1/4}\\
 	&\approx& v t \left(1-\frac{\lambda t^2}{2}\right),\nonumber
\end{eqnarray}
with $\lambda > 0$ in both cases. The superlinear case has $\epsilon^{(3)}(0)\times \dot{\epsilon}(0)>0$, while the sublinear case has $\epsilon^{(3)}(0)\times \dot{\epsilon}(0)<0$. Throughout this manuscript, we denote the first and second time derivatives using dots, while we denote the third time derivative with the superscript ``(3)''. The functions plotted in Fig.~\ref{Fig:Schematic}(c) have the additional property of being antisymmetric about the point $t=0$ and therefore contain only odd powers of $t$.

\section{Dykhne-Davis-Pechukas formula}
\label{Sec:DDP}

The DDP formula \cite{Dykhne,Davis} (see also Ref.~\cite{LandauLifshitz}) is a powerful tool to investigate LZSM-type problems with general functional forms of the parameters $v(t)$ and $\Delta(t)$. It was derived with rigorous foundation in the adiabatic limit, but it has proved to be a good approach to obtain rather accurate results even away from the adiabatic limit.

A calculation utilizing the DDP formula typically proceeds as follows: for the time evolution from $t\rightarrow -\infty$ to $t\rightarrow \infty$ with general functions $\epsilon(t)$ and $\Delta(t)$, one first finds the zeros of the function
\begin{equation}
\mathcal{E}(t) = \sqrt{\epsilon^2(t)+\Delta^2(t)}
\end{equation}
in the complex plane. In other words, one finds the values of $t$ that satisfy the condition
\begin{equation}\label{EquationForZeroPoints}
\mathcal{E}(t) = 0,
\end{equation}
treating $t$ as a complex variable. For the DDP approach to be justified, the following assumptions are made:

(1) the function $\mathcal{E}(t)$ is analytic at least in a region that contains the real axis and relevant zeros, 

(2) there are no zeros on the real $t$ axis, i.e.~the functions $\epsilon(t)$ and $\Delta(t)$ do not vanish simultaneously at any point during the parameter variation, and 

(3) no two zeros are extremely close or coincide with each other.

One then identifies all the zeros of $\mathcal{E}(t)$ that have positive imaginary parts. In general, there will be multiple such zeros $t_c^k$, where the superscript $k$ is a label for the different zeros ($k=1, 2, ..., n$). In this case, the \textit{generalized DDP formula} can be used to calculate the transition probability \cite{Vitanov}
\begin{equation}\label{Eq:GeneralizedDDP}
	P\approx\left|\sum_{k=1}^n\Gamma_k e^{iD(t_c^k)}\right|^2,
\end{equation}
where
\begin{equation}
	\Gamma_k=4i\lim_{t\to t_c^k}(t-t_c^k)\frac{\dot{\epsilon}(t)\Delta(t)-\epsilon(t)\dot{\Delta}(t)}{2\mathcal{E}^2(t)},
\end{equation}
and
\begin{equation}
D(t) = \int_0^t \mathcal{E}(s) ds.
\end{equation}
Note that $|\Gamma_k|=1$ for functions that satisfy the required conditions for the DDP approach. If there is only one zero of $\mathcal{E}(t)$ with a positive imaginary part, i.e.~$n=1$, Eq.~(\ref{Eq:GeneralizedDDP}) is considerably simplified. Even if there are multiple zeros, one can obtain a simple approximation for Eq.~(\ref{Eq:GeneralizedDDP}) by keeping only the contribution from the zero that has the smallest positive imaginary part, i.e.~the zero that is closest to the real axis in the upper half of the complex plane. This approximation can be justified by the reasonable argument that as we move farther away from the real axis we can intuitively expect that the integrals $D(t_c^k)$, including their positive imaginary parts, will grow, which means that the contributions of far-away zeros to Eq.~(\ref{Eq:GeneralizedDDP}) will be exponentially small and can be neglected. One then obtains the standard DDP formula for the transition probability,
\begin{equation}
P = e^{-2 {\rm Im} D(t_c)}.
\label{Eq:DDP}
\end{equation}
We note here that the generalized DDP formula was put on rigorous foundations in Refs.~\cite{Suominen1991,Suominen1992a,Suominen1992b}, although it was mentioned in Ref.~\cite{Davis} as a possible generalization for the single-zero formula of Eq.~(\ref{Eq:DDP}).

In the simple case $\epsilon(t)=vt$ and $\Delta(t)=\Delta$, i.e.~the original LZSM problem, we easily find that there is only one relevant zero, namely $t_c=i\Delta/v$, which then gives the simple elliptical integral
\begin{eqnarray}
D(t_c) & = & i\int_0^{\Delta/v}\sqrt{\Delta^2-(vs)^2}ds \nonumber \\
& = & \frac{i\Delta^2}{v} \int_0^1\sqrt{1-x^2}dx 
= \frac{i\pi\Delta^2}{4v},
\label{Eq:DtcForLinearLZSM}
\end{eqnarray}
leading to the LZSM formula [Eq.~(\ref{Eq:P_LZ})].

\section{Previous work on the nonlinear LZSM problem}
\label{Sec:PreviousWork}

We now briefly review previous studies of the nonlinear LZSM problem in the literature.

\subsection{Perturbative nonlinearities}

Here, the nonlinearities appear as perturbative corrections, and the transition probability $P$ deviates little from the LZSM formula. These corrections are in fact of principal interest for our present study. With such perturbations, appreciable deviations from the LZSM formula appear only for sufficiently large values of the nonlinearity parameter. By reducing the nonlinearity to zero, one can approach and recover the LZSM formula.

Based on the DDP formalism, the superlinear (\ref{superlinear}) and sublinear (\ref{sublinear}) sweep functions were considered in Ref.~\cite{Vitanov}. In the superlinear and sublinear cases, the nonlinearity, respectively, decreases and increases the transition probability $P$ with respect to the LZSM formula \cite{Vitanov}.

In Ref.~\cite{Vaezi}, the authors studied situations where the Hamiltonian of a gapless physical system, such as graphene and 1- and 2-dimensional $p$-wave superconductors, cannot be linearized even in the vicinity of the crossing point and quadratic corrections are crucial. Similarly, linearizing the spectrum in the vicinity of the crossing point is insufficient to describe the probability of the topological transition in a 2-dimensional electron gas subject to an in-plane magnetic field and in the presence of spin-orbit coupling \cite{Malla2019}. In this context it is worth mentioning the related problem in which a physical system (for example, single electrons tunneling between semiconductor quantum dots) possesses more than two states and experiences nonadiabatic transitions among these states, which is referred to as the multistate LZSM problem \cite{Malla2021}.

\subsection{Essential nonlinearities}

In this case, the problem parameters do not simply have a small correction term added to the linear LZSM problem, e.g.~having a bias term $\epsilon = v t + \delta\epsilon(t)$ where $\delta\epsilon(t)$ is a small term that leads to small corrections in various physical quantities. Instead, the problem parameters deviate essentially from the linear case. Essential nonlinearities in the bias term of power-law form
\begin{equation}
\epsilon (t) = \beta ^{N+1} t^{N},
\end{equation}
(with $\beta $ having frequency units) were studied in detail in Refs.~\cite{Vitanov} and \cite{Lehto2012} for odd and even $N$, respectively, both analytically with the generalized DDP formula and by numerically solving the Schr\"odinger equation.

The case with a general value of $N$, not necessarily an integer, was treated in Ref.~\cite{GaraninPRB}. Note that by treating $N$ as a real number and taking the limit $N\to 1$ the nonlinearity becomes perturbative.

The special case of $N=2$ is called the parabolic model~\cite{Lehto2012}. In general, this model is characterized by the sweep function
\begin{equation}
\epsilon (t)=\epsilon _{0}+\alpha t^{2}.
\end{equation}
For $\epsilon _{0} \times \alpha < 0$, we obtain a double passage of the avoided crossing. Since the two passages do not occur at $t=0$, each passage can be approximated as a linear one with the parameters taken from their values at the points of the energy quasi-crossing \cite{Shimshoni}. Hence, this situation is, in general, not a case of an essential nonlinearity. It becomes an essentially non-linear model at $\epsilon_{0}=0$.

Essential nonlinearities also include non-analytical models, e.g.~models with sweep functions that are non-analytical at $t=0$ \cite{GaraninPRB,Damski}.

\subsection{Exactly solvable problems with nonlinearities}

Several models assume analytic solutions, typically because the functional forms of $\epsilon(t)$ and $\Delta(t)$ allow an analytic integration of the Schr\"odinger equation. Such models can be used as test-beds for different computational and analytic approaches.

One such case is the so-called Allen-Eberly-Hioe model \cite{Allen,Hioe}, which is a special case of the Demkov-Kunike model \cite{Demkov} (see also Ref.~\cite{Suominen1992a}):
\begin{equation}
	\epsilon (t)=2B\tanh (t/T),\text{ \ \ }\Delta (t)=\frac{2A}{\cosh (t/T)}.
\end{equation}
This choice of functions results in the transition probability, i.e.~the probability to remain in the same diabatic state,
\begin{equation}
	P=\frac{\cosh ^{2}\left( \pi \sqrt{B^{2}-A^{2}}T\right) }{\cosh
		^{2}\left( \pi BT\right) }  \label{cosh_cosh}
\end{equation}
(see Ref.~\cite{Suominen1992a}). This probability is equal to unity ($P=1$) at $A=0$ and decreases with increasing $A$ for $A<B$. When $A>B$, the probability is an oscillatory function of $\sqrt{A^2-B^2}T$. Another member of this class, with the very same result, i.e.~Eq.~(\ref{cosh_cosh}), is the case of sweeping through an avoided crossing with the tangent function 
\begin{equation}
	\epsilon (t)=2B\tan (t/T),\text{ \ \ }\Delta (t)=2A, \,\,\,\,
	-\frac{\pi T}{2}\leq t \leq \frac{\pi T}{2}.
\end{equation}
As can be expected, in the limit $B/A\to\infty$, the nonlinearity is weak, leading as a result to a small correction to the LZSM formula. This correction can also be obtained via the DDP formula \cite{Vitanov}.

The Rosen-Zener model,
\begin{equation}
	\epsilon (t)=2a,\text{ \ \ }\Delta (t)=\frac{2b}{\cosh(t/T)},
\end{equation}
was introduced in Ref.~\cite{Rosen} (see also Refs.~\cite{Suominen1992a,Lehto2016}). Because there is no avoided-level crossing in this problem, there is no LZSM limit. The probability to remain in the same diabatic state is given by \cite{Suominen1992a}
\begin{equation}
	P = 1 - \frac{\sin ^{2}(\pi bT)}{\cosh^2(\pi aT)},
\end{equation}
which oscillates as a function of $b$ and is equal to unity at integer values of $b T$.

\subsection{Nonlinear LZSM problem in Bose-Einstein condensates}

A special case of the nonlinear LZSM problem relates to a two-level system where the level energies depend on the occupation probabilities of the two levels. This situation can arise in the mean-field treatment of a many body system where the particles predominantly occupy two quantum states and the inter-particle interaction energy depends on the states of the particles \cite{Liu,Ishkhanyan,Carles,Wu}. Such systems can be described by the Gross-Pitaevskii equation, which plays a similar role as the Schr\"odinger equation but can be nonlinear in the probability amplitudes \cite{Trimborn,Leggett}. Even if the external system parameters vary linearly, the problem can be equivalent to a nonlinear one because of the nonlinearity that appears implicitly through the interaction term, keeping in mind that the form of the nonlinearity is determined only when the problem is solved. Such a system could be experimentally realized in several ways within the context of Bose-Einstein condensate \cite{Zhang}.

\subsection{Reverse engineering and transitionless driving}

The idea of reverse engineering can be formulated as finding a Hamiltonian $\widetilde{H}(t)$ that generates a given dynamics, e.g.~a certain evolution in the basis of the instantaneous eigenstates of a given Hamiltonian $H(t)$ \cite{Berry} (see also the review \cite{GueryOdelin}). One example is the inverse LZSM problem, formulated as finding the bias $\epsilon (t)$ that results in any required time dependence of the level populations. This problem was solved in Ref.~\cite{GaraninEPL} (see also Ref.~\cite{Shevchenko2012}). Another problem related to the reverse engineering approach is the problem of transitionless quantum driving, which is analogous to reflectionless potentials. This problem was studied both theoretically \cite{Berry,Villazon} and experimentally \cite{Bason,Xu}. Importantly, in our context, the linear driving requires a nonlinear correction in $\widetilde{H}(t)$ so that the resulting dynamics becomes transitionless \cite{Berry}.

\section{Perturbative nonlinearity}
\label{Sec:PerturbativeNonlinearity}

In this section, we derive analytical perturbative formulae for the corrections to the transition probability caused by small quadratic and cubic nonlinearities. We also present results of numerical calculations and compare them with the perturbative formulae. Finally, we comment on the application of the DDP approach to the double-passage problem with a linear-plus-quadratic sweep function.

\subsection{Linear-plus-quadratic sweep function}
\label{Sec:PerturbativeDDPQuadratic}

\begin{figure}[h]
\includegraphics[width=8.5cm]{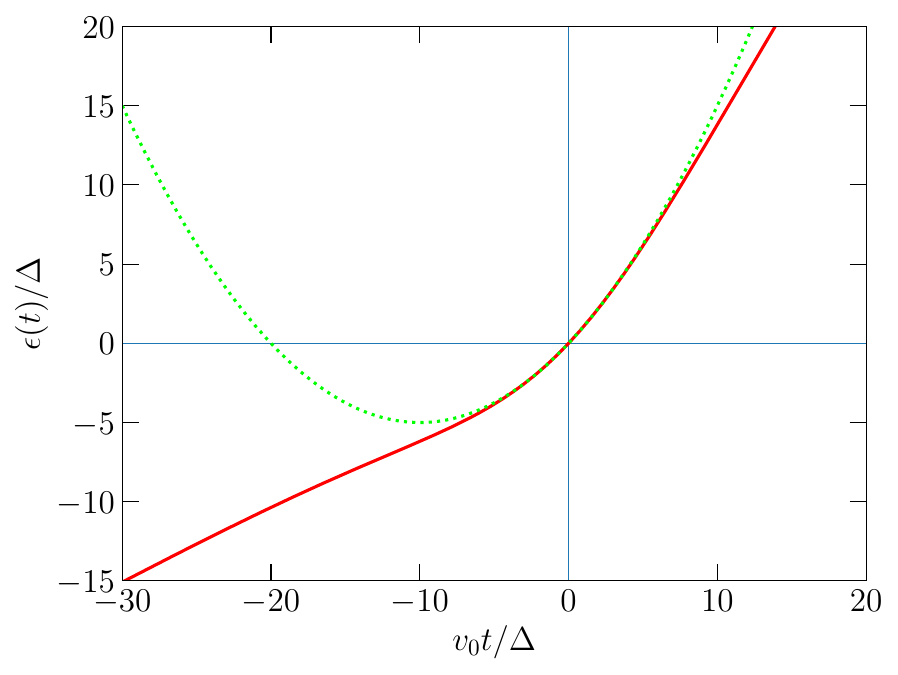}
\caption{Two possible nonlinear sweep functions that contain quadratic terms: the solid red line is the function $\epsilon(t) = v_0 t [1 + 0.5 \tanh (v_0t/10\Delta)]$, while the dotted green line is the function $\epsilon(t) = v_0 t + ( 0.05 v_0^2 / \Delta) t^2$. These two functions have the same first and second derivatives at $t=0$. However, when we consider the behavior from $t \rightarrow -\infty$ to $t \rightarrow \infty$, the two functions describe qualitatively different LZSM problems: the former describes a single-passage problem, while the latter describes a double-passage problem.}
\label{Fig:TwoSweepFunctions}
\end{figure}

We start by taking the linear LZSM problem and adding the lowest-order correction to the sweep function $\epsilon(t)$:
\begin{equation}\label{QuadraticCorrection}
\epsilon(t) = \dot{\epsilon}(0)t+\frac{1}{2}\,\ddot{\epsilon}(0)t^2 = 
 v_0 t + \frac{\chi_2}{2\Delta} (v_0 t)^2,
\end{equation} 
where we have introduced the parameter $v_0$ to denote the sweep rate at the crossing point, i.e.~$v_0=\dot{\epsilon}(0)$. To simplify the notation, we will sometimes find it convenient to express the sweep function in Eq.~(\ref{QuadraticCorrection}) as $\epsilon(t) = v_0 t + v_1 t^2$, hence defining $v_1=\chi_2 v_0^2/(2\Delta)$. At first sight, this sweep function seems to be the natural one containing the lowest-order nonlinearity, namely the quadratic term. However, this function is obviously not a good choice for our purposes, because $\epsilon(t)$ approaches the same value (either $+\infty$ or $-\infty$) when $t\rightarrow -\infty$ and when $t\rightarrow \infty$, as shown in Fig.~\ref{Fig:TwoSweepFunctions}. As a result, we would not simply obtain perturbative corrections to the LZSM formula. Instead, we would encounter two crossings. One consequence of this situation is that in both the adiabatic and the fast limits the system is expected to return to its initial state at the final time. However, we are interested in the effect of the nonlinearity on the single-passage problem. We therefore look for a function $v(t)$ that varies almost linearly in the vicinity of $t=0$ but becomes constant away from $t=0$. In our numerical simulations of the dynamics, we shall use the tanh function, which exhibits the above-described dependence on $t$. Specifically, we use the sweep function
\begin{equation}
\epsilon(t) = v_0 t \times \left( 1 + \alpha \tanh \frac{t}{T} \right),
\label{Eq:VariableRateSweepFunction}
\end{equation}
where $\alpha$ quantifies the total variation in the sweep rate over time, and $T$ is the duration over which the temporal variation of $v(t)$ continues. If $T$ is much larger than the crossing duration [Eq.~(\ref{Eq:CrossingTime})], one can expect that the tanh function has the intended effect of providing a quadratic term in $\epsilon(t)$ throughout the most consequential time for the LZSM transition dynamics without creating a double-passage situation. To avoid the problem of a double crossing, it is obviously important that the value of $v(t)$ does not change sign, which requires us to take $|\alpha|<1$.

As explained in the previous paragraph, using Eq.~(\ref{QuadraticCorrection}) results in a qualitatively different problem from the one that we would like to study. However, the drastic deviation from the linear case, namely the turnaround in $\epsilon(t)$, occurs for large negative values of the time variable $t$. For small values of $|t|$, the quadratic term in $\epsilon(t)$ does indeed appear to be the small perturbation that we would like to include. We therefore perform a perturbative calculation where we follow the steps for evaluating the DDP formula to determine the transition probability in the presence of this term, being careful to include only the small corrections that vanish in the linear, single-passage problem.

As explained in more detail in Appendix \ref{App:QuadraticNonlinearity}, the zero in the DDP calculation that used to be at $t_c=i\Delta/v$ is shifted because of the nonlinear term to 
\begin{equation}
	t'_c \approx \frac{\Delta}{v_0} \left\{\frac{\chi_2}{2} + i\left(1-\frac{\chi_2^2}{2}\right)\right\},
\end{equation}
this approximation being valid when $\chi_2$ is small:
\begin{equation}
\chi_2 \ll 1.
\label{Eq:SmallChiCondition}
\end{equation}

After the substitution $x=(t'_c-s)/t'_c$, the integral $D(t'_c)=\int_0^{t'_c}{\mathcal{E}(s)ds}$ takes the form
\begin{equation}
D(t'_c) = -i v_0 (t'_c)^2\int_0^1 Q(x) dx,
\end{equation}
where
\begin{equation}
Q(x) = {\sqrt{p_2(x)}\sqrt{1+2\gamma\frac{p_3(x)}{p_2(x)} + \gamma^2\frac{p_4(x)}{p_2(x)}}},
\end{equation}
$\gamma=\chi_2 v_0 t'_c/(2\Delta)$, $|\gamma|\sim \chi_2\ll 1$, $p_2(x)=2x-x^2$, $p_3(x)=3x-3x^2+x^3$, $p_4(x)=4x-6x^2+4x^3-x^4$. Using the Taylor expansion of the square root and holding terms up to $\gamma^2$ we obtain 
\begin{equation}
D(t'_c)\approx \frac{1}{3}\frac{\Delta^2}{v_0}\chi_2 + i\frac{\pi}{4}\frac{\Delta^2}{v_0}\left(1-\frac{3\chi_2^2}{8}\right).
\label{Eq:DDPIntegralQuadratic}
\end{equation}
It is worth noting that two things go wrong if the condition (\ref{Eq:SmallChiCondition}) is not satisfied: the approximate expression for $t'_c$ and the Taylor expansion in the integral of $D(t'_c)$ both break down outside the small $\chi_2$ regime.

Using the value of $D(t'_c)$ given in Eq.~(\ref{Eq:DDPIntegralQuadratic}), the transition probability including the lowest-order correction reads
\begin{equation}\label{Eq:LZFormulaQuadratic}
	P \approx \exp \left\{-2\pi\delta \left(1 - \frac{3\chi_2^2}{8} \right) \right\},
\end{equation}
with the adiabaticity parameter $\delta$ defined in Eq.~(\ref{Eq:AdiabaticityParameter}) with $v=\dot{\epsilon}(0)=v_0$.
The first term inside the exponential is the one that gives the standard LZSM formula (\ref{Eq:P_LZ}), while the second term represents the correction arising from the nonlinearity in the problem. Since our perturbative DDP calculation is valid when $\chi_2 \ll 1$, we can also write the alternative approximate expression
\begin{equation}
P \approx P_{\rm LZSM}\left(1 + \frac{3}{4}\pi\delta \chi_2^2 \right).
\label{Eq:CorrectionToLZFormula}
\end{equation}
Interestingly, this approximation remains well behaved and a good approximation even if we take the adiabatic limit with $\chi_2 \gtrsim 1$, as we shall see below. If we vary $v_0$ while keeping the ratio $v_1/(v_0\Delta)$ fixed, the function on the right-hand side of Eq.~(\ref{Eq:CorrectionToLZFormula}) has a peak whose highest value of $(27e^{-3}/\pi)\times v_1^2/v_0^3$ is located at $v_0/\Delta^2=\pi/6$.

It is interesting that the correction in Eq.~(\ref{Eq:LZFormulaQuadratic}) skips one power, i.e.~there is no term that is linear in $\chi_2$. Since these extra terms inside the exponential describe the correction to the LZSM formula, skipping the lowest possible power means that the correction will be quite small, as we shall see more clearly when we present the numerical results in Sec.~\ref{Sec:NumericalResults}. It is also interesting that $P$ depends on the magnitude but not the sign of $\chi_2$, i.e.~it is independent of whether the nonlinearity corresponds to a sweep that is speeding up or slowing down during the crossing.

It is useful at this point to consider the physical meaning of the condition $\chi_2\ll 1$ needed for the validity of our derivations above. In the slow-passage regime ($v_0\ll\Delta^2$), considering that the crossing duration is given by $\tau_{\rm LZSM}\sim\Delta/(2v_0)$ [see Eq.~(\ref{Eq:CrossingTime})], the condition $\chi_2\ll 1$ can be understood as the condition that the quadratic term in $\epsilon(t)$ remains much smaller than the linear term throughout the duration of the crossing process. This condition makes sense. In the fast-passage regime ($v_0\gg\Delta^2$), $\tau_{\rm LZSM}\sim 1/\sqrt{v_0}$, and the condition $\chi_2\ll 1$ does not guarantee that the quadratic term in $\epsilon(t)$ is small compared to the linear term up to times $t\sim\pm\tau_{\rm LZSM}$. Indeed, our numerical simulations show that the DDP approach is valid only under a stricter weak-nonlinearity condition in the fast-passage regime, as we shall see in Sec.~\ref{Sec:NumericalResults}.

\subsection{Linear-plus-cubic sweep function}
\label{Sec:PerturbativeDDPCubic}

Now we consider the case where the sweep function does not contain a quadratic term, and the leading-order nonlinearity (at the crossing point) is cubic:
\begin{equation}
	\epsilon(t) = \dot{\epsilon}(0)t+\frac{1}{3!}\,\epsilon^{(3)}(0)t^3
 = v_0 t + \frac{\chi_3}{3!\Delta^2} (v_0 t)^3.
	\label{FirstThird}
\end{equation}
This function describes a perturbative nonlinearity if the second term is small compared to the first one throughout the crossing duration. Assuming that $\chi_3$ is small, we can follow steps similar to those of Sec.~\ref{Sec:PerturbativeDDPQuadratic} and obtain
\begin{equation}
	t'_c=i\frac{\Delta}{v_0}\left(1+\frac{\chi_3}{6}\right)
\end{equation}
and 
\begin{equation}
	\text{Im}D(t'_c)=\frac{\pi}{4}\frac{\Delta^2}{v_0}\left(1+\frac{\chi_3}{8}\right).
\end{equation}
Then we obtain for the transition probability 
\begin{eqnarray}\label{Eq:LZFormulaCubic}
	P & \approx & \exp \left\{ -2\pi\delta \left(1+\frac{\chi_3}{8}\right) \right\} \nonumber \\ 
	& \approx & P_{\rm LZSM} \left(1-\frac{\pi\delta \chi_3}{4}\right),
\end{eqnarray}
$\delta$ being the adiabaticity parameter defined in Eq.~(\ref{Eq:AdiabaticityParameter}) with $v=v_0$.

Importantly, in contrast to the case of a quadratic nonlinearity, the correction from the cubic nonlinearity is linear in $\chi_3$. Note that Eq.~(\ref{Eq:LZFormulaQuadratic}) for the quadratic correction and the first row of Eq.~(\ref{Eq:LZFormulaCubic}) can be expressed in the form $P\approx\exp\{-2\pi\delta(1+\mu)\}$, with 
$\mu=-3\chi_2^2/8$ and $\chi_3/8$, respectively. Furthermore, we can combine the perturbative corrections in Eqs.~(\ref{Eq:LZFormulaQuadratic}) and (\ref{Eq:LZFormulaCubic}) and obtain the unified formula
\begin{equation}
P \approx \exp \left\{-2\pi\delta \left(1 - \frac{3\chi_2^2}{8} + \frac{\chi_3}{8} \right) \right\}.
\label{Eq:LZFormulaQuadraticCubic}
\end{equation}
In Sec.~\ref{Sec:VariableDelta} and Appendix \ref{App:TimeDependentGap} we discuss a case where this formula is crucial and proves to be a good approximation.

\subsection{Numerical results}
\label{Sec:NumericalResults}

To demonstrate the validity and limitations of the approximate formulae derived in the previous subsections, we performed essentially exact numerical simulations of the dynamics by solving the time-dependent Schr\"odinger equation and compared the results to those predicted by our approximate formulae.

In our calculations for the case of a quadratic nonlinearity we use the tanh-function-based sweep function [Eq.~(\ref{Eq:VariableRateSweepFunction})]. For the case of a cubic nonlinearity, we simply use Eq.~(\ref{Eq:NonlinearEpsilonExpansion}) with $\chi_2=0$, since the cubic term does not lead to the same complications as the quadratic term. For both cases, as well as in other similar situations that we consider below, we vary the sweep rate at the crossing point [i.e.~$\dot{\epsilon}(0)$ or $v_0$] and use it for the $x$ axis when we plot the calculation results.

To confirm that our simulations using Eq.~(\ref{Eq:VariableRateSweepFunction}) correctly represent the desired nonlinearity, we perform multiple calculations with different settings that are expected to produce the same results, for example varying $\alpha$ and $T$ in Eq.~(\ref{Eq:VariableRateSweepFunction}) while keeping the ratio $\alpha/(T\Delta)$ fixed. Any deviation between the results of these alternative calculations would be an indication that our parameters are not suitable to produce the sought physical results. Similarly, in our numerical calculations, we run multiple calculations where we vary the initial and final times, which will necessarily be finite in a numerical simulation, to make sure that we are obtaining the correct asymptotic values for the transition probability.

\begin{figure}[h]
\includegraphics[width=8.5cm]{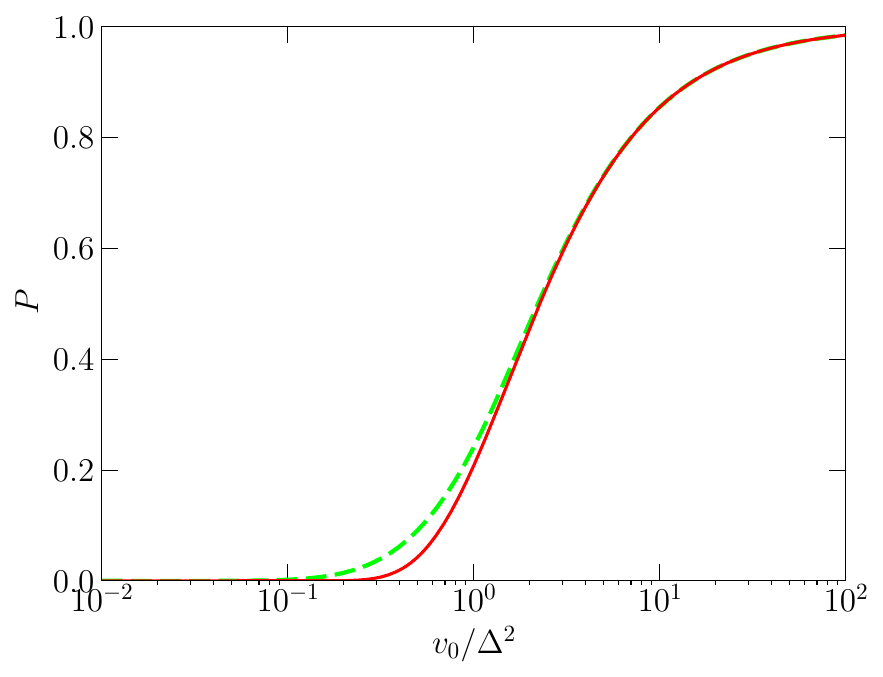}
\caption{Probability that the system makes a transition between the ground and excited states as a result of traversing the avoided crossing. The parameter $v_0$ is the sweep rate at the crossing point, where $\epsilon(t)=0$. All our simulations of the LZSM problem with weak nonlinearities produce probability functions that look generally similar to the two shown here. The solid red line is the transition probability $P_{\rm LZSM}$ for the linear LZSM problem. The dashed green line corresponds to the sweep function in Eq.~(\ref{Eq:VariableRateSweepFunction}) with $\alpha=0.8$ and $T=3/\Delta$, which will be analyzed further in Fig.~\ref{Fig:ProbabilityDeviationStrong} below.}
\label{Fig:Probability}
\end{figure}

\begin{figure}[h]
\includegraphics[width=8.5cm]{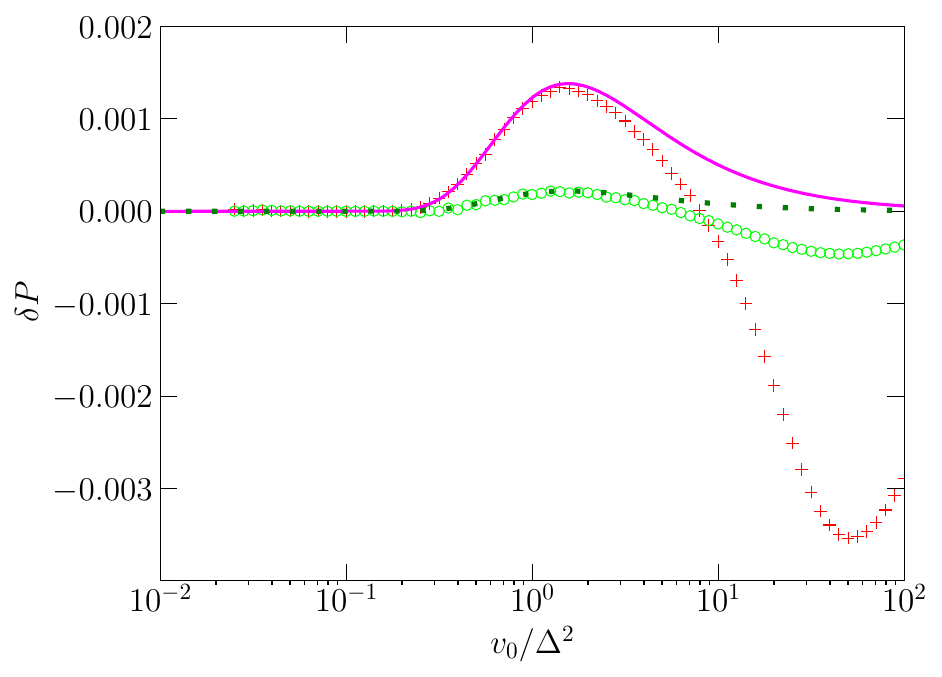}
\caption{Deviation of the transition probability in the nonlinear case from the linear-case probability $P_{\rm LZSM}$. The data plotted using the various symbols are obtained by numerically solving the Schr\"odinger equation and can be considered exact, while the data plotted as lines are obtained using the DDP approach. For the red + symbols and green circles, we use the sweep function described by Eq.~(\ref{Eq:VariableRateSweepFunction}) with $T=10\Delta/v_0$. The red + symbols correspond to $\alpha=0.5$ ($\chi_2=0.1$). The green circles correspond to $\alpha=0.2$ ($\chi_2=0.04$). The solid magenta line and dotted green line show the deviation $\delta P$ calculated based on Eq.~(\ref{Eq:LZFormulaQuadratic}) with $\chi_2=0.1$ and 0.04, respectively. The DDP formula accurately reproduces $\delta P$ at small values of $v_0/\Delta^2$, but clear disagreement is observed at large values of $v_0/\Delta^2$. As explained in the text, in this regime the quadratic term is not a small perturbation throughout the crossing duration.}
\label{Fig:ProbabilityDeviationFixedNonlinearityWeak}
\end{figure}

\begin{figure}[h]
\includegraphics[width=8.5cm]{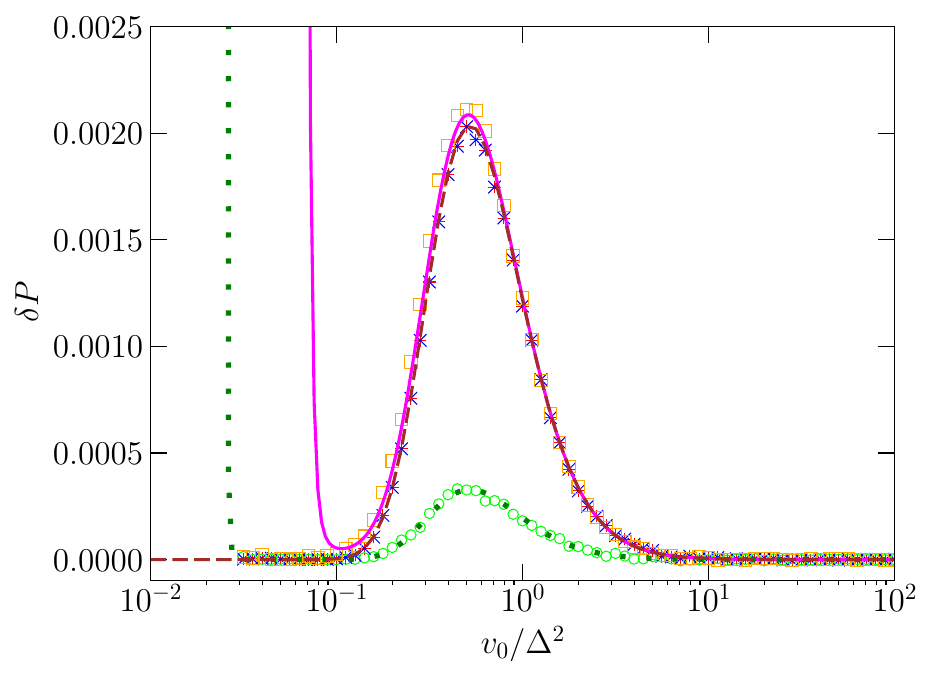}
\caption{Same as in Fig.~\ref{Fig:ProbabilityDeviationFixedNonlinearityWeak}, but keeping ratio $v_1/(v_0\Delta)$ fixed instead of keeping $\chi_2$ fixed. For all the data shown by symbols in this figure, we set $T=10/\Delta$. The red + symbols correspond to $\alpha=0.5$. The blue $\times$ symbols correspond to $\alpha=-0.5$. The green circles correspond to $\alpha=0.2$. The orange squares correspond to a time-independent $v$ and time-dependent $\Delta$ given by $\Delta_0\left(1+0.5\tanh(\Delta_0 t/10) \right)$. The solid magenta line and dotted green line show the deviation $\delta P$ calculated based on Eq.~(\ref{Eq:LZFormulaQuadratic}) with $v_1/(v_0\Delta)=0.05$ and 0.02, respectively. The magenta line also corresponds to Eq.~(\ref{Eq:LZFormulaVariableDelta}) with $\Delta'=0.05$. The approximate theoretical formula breaks down at small values of $v_0$ because the condition $\chi_2\ll 1$ is no longer satisfied in that regime. Importantly, the theoretical formula fits the simulation results very well in the region of largest nonlinearity-induced correction. The dashed brown line shows the results of a DDP calculation with all the steps performed numerically for the case of the parameters that gave the + symbols. This calculation gives good agreement with the + symbols everywhere. Comparing the + and $\times$ symbols, we see that the cases $\alpha=0.5$ and $\alpha=-0.5$ give identical results, as expected.}
\label{Fig:ProbabilityDeviationVariableNonlinearityWeak}
\end{figure}

\begin{figure}[t]
\includegraphics[width=8.5cm]{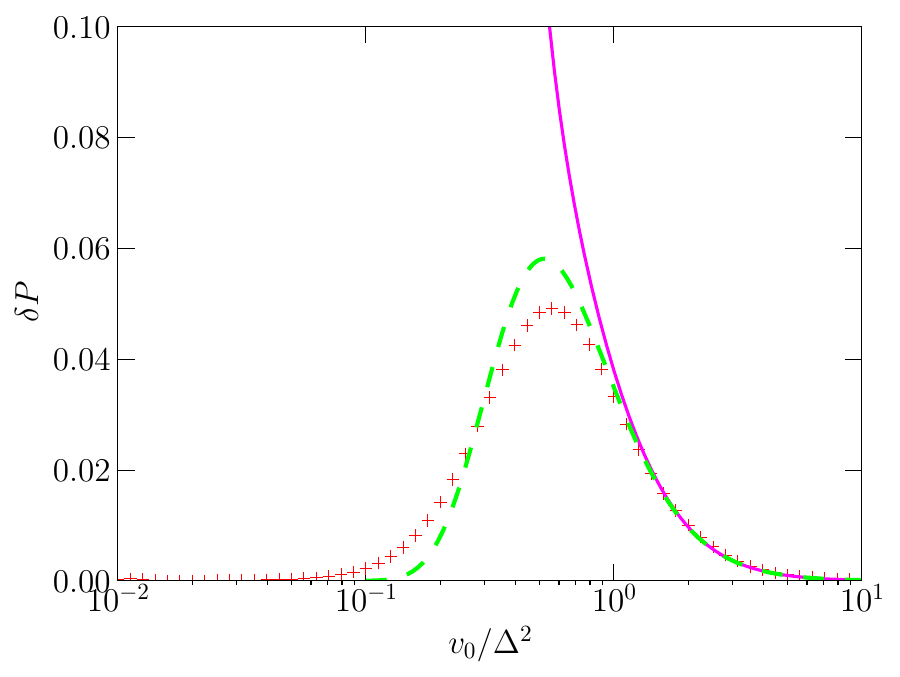}
\caption{Same as in Fig.~\ref{Fig:ProbabilityDeviationVariableNonlinearityWeak}, but for stronger nonlinearity. Now we take $\alpha=0.8$ and $T=3/\Delta$. The red + symbols correspond to the solution of the Schr\"odinger equation. The solid magenta line and dashed green line are, respectively, given by Eqs.~(\ref{Eq:LZFormulaQuadratic}) and (\ref{Eq:CorrectionToLZFormula}) with $v_1/(v_0\Delta)=4/15$. The perturbative formula [Eq.~(\ref{Eq:LZFormulaQuadratic})] clearly fails for small values of $v_0/\Delta^2$, while Eq.~(\ref{Eq:CorrectionToLZFormula}), which we obtained as an approximation for Eq.~(\ref{Eq:LZFormulaQuadratic}) in the small $v_1\Delta/v_0^2$ limit, turns out to be a reasonably good approximation for all values of $v_0/\Delta^2$.}
\label{Fig:ProbabilityDeviationStrong}
\end{figure}

\begin{figure}[h]
\includegraphics[width=8.5cm]{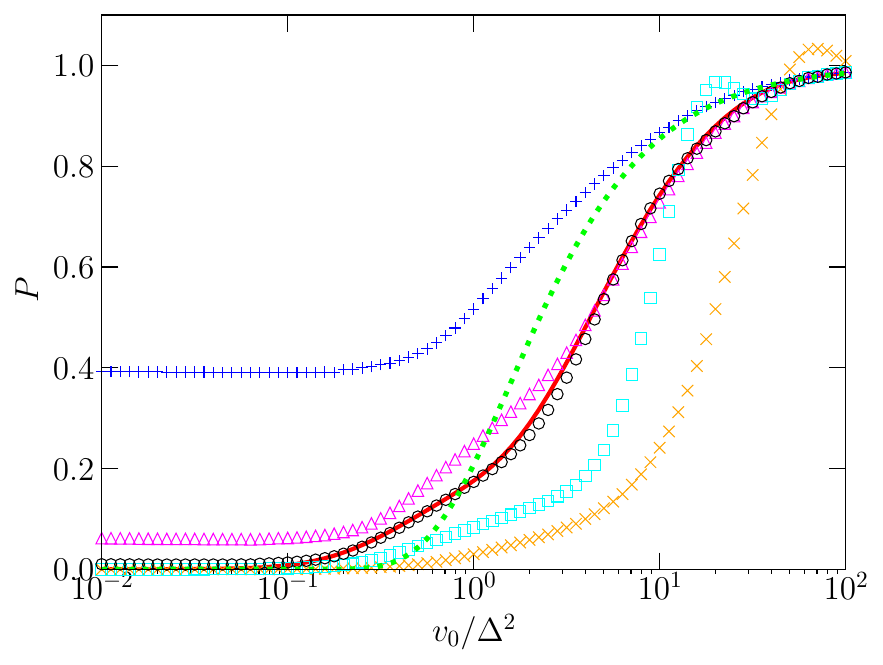}
\caption{Transition probability $P$ as a function of $v_0/\Delta^2$ for the case of very strong nonlinearity: $\alpha=0.8$ and $T=0.3/\Delta$. The dotted green line is $P_{\rm LZSM}$ and serves as a reference. The solid red line is obtained by numerical integration of the time-dependent Schr\"odinger equation. The other data points are obtained by numerical evaluation of the transition probability following the DDP approach. The blue + symbols, orange $\times$ symbols, magenta triangles, cyan squares and black circles correspond, respectively, to keeping 1, 2, 3, 4 and 5 zeros of $\mathcal{E}(t)$ in the DDP calculation.}
\label{Fig:ProbabilityDeviationVeryStrong}
\end{figure}

\begin{figure}[h]
\includegraphics[width=8.5cm]{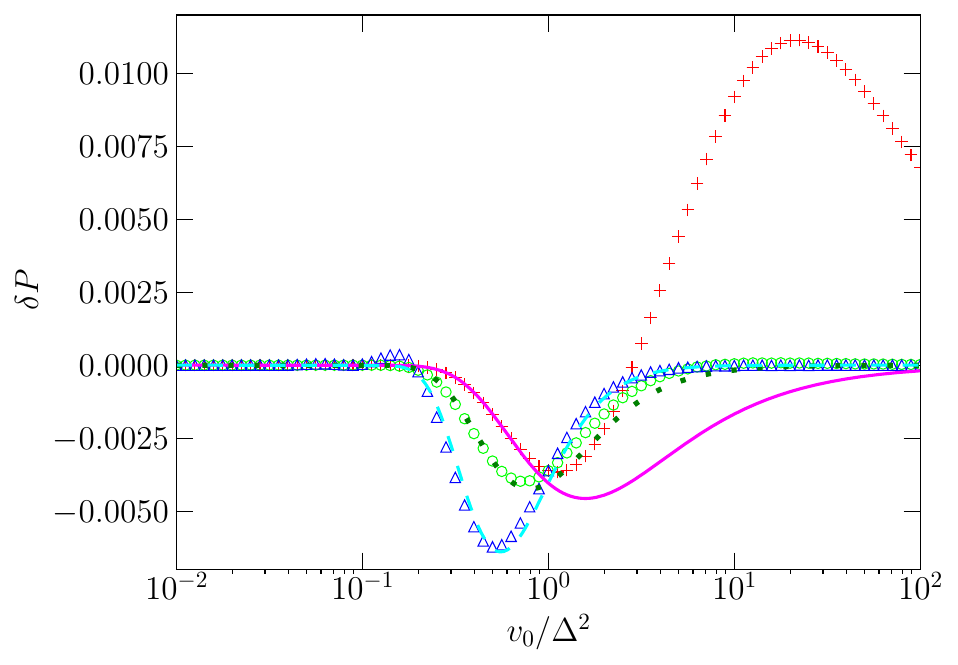}
\caption{Probability difference $\delta P$ as a function of $v_0/\Delta^2$ for the sweep function $\epsilon(t)=v_0t + \chi_3 (v_0 t)^3/(6\Delta^2)$. The red + symbols, green circles and blue triangles correspond, respectively, to $\chi_3=0.1$, $0.1\Delta^2/v_0$ and $0.1\Delta^4/v_0^2$. The solid magenta, dotted green and dashed cyan lines are obtained using Eq.~(\ref{Eq:LZFormulaCubic}) with $\chi_3$ values that correspond, respectively, to the red + symbols, green circles and blue triangles. Keeping $\chi_3$ fixed leads to large deviations between the perturbative DDP formula and exact results in the fast-passage regime, while having a value of $\chi_3$ that decreases with increasing $v_0/\Delta^2$ leads to better agreement between the approximate formula and exact transition probability.}
\label{Fig:ProbabilityDeviationCubic}
\end{figure}

Since here we consider small deviations from the linear LZSM problem, the transition probability as a function of the sweep rate for all our data sets look generally like the curves in Fig.~\ref{Fig:Probability}. To investigate the corrections to the LZSM probability, we plot the difference between the transition probability $P$ for a given set of parameters and the LZSM probability in the linear case:
\begin{equation}
\delta P = P-P_{\rm LZSM}. 
\end{equation}
Figures \ref{Fig:ProbabilityDeviationFixedNonlinearityWeak} -- \ref{Fig:ProbabilityDeviationSinhVariableNonlinearity} show the probability $P$ or the deviation $\delta P$, depending on which one is more informative, for a few different cases of LZSM problems with perturbative nonlinearities.

First we fix the quadratic nonlinearity coefficient $\chi_2$ and vary $v_0/\Delta^2$. In other words, we use Eq.~(\ref{Eq:VariableRateSweepFunction}) with $\alpha/T=\chi_2 v_0/(2\Delta)$. The results are shown in Fig.~\ref{Fig:ProbabilityDeviationFixedNonlinearityWeak}. The perturbative DDP formula [Eq.~(\ref{Eq:LZFormulaQuadratic})] agrees with the exact solution in the slow-passage regime but becomes invalid in the fast-passage regime. As explained in Sec.~\ref{Sec:PerturbativeDDPQuadratic}, this breakdown can be attributed to the fact that a fixed $\chi_2$ and an increasingly large $v_0/\Delta^2$ lead to a situation where the nonlinear term is not small throughout the crossing duration. We note here that the exact simulation exhibits a sign reversal in $\delta P$, whereas Eq.~(\ref{Eq:LZFormulaQuadratic}) always gives a positive value of $\delta P$. To investigate the situation where the nonlinear term is small in the fast-passage limit, we perform additional simulations in which we keep the ratio $v_1/(v_0\Delta)$, i.e.~$\chi_2 v_0/(2\Delta^2)$, fixed and small instead of keeping $\chi_2$ fixed. As can be seen in Fig.~\ref{Fig:ProbabilityDeviationVariableNonlinearityWeak}, the perturbative DDP formula now agrees well with our numerical simulation results for a small nonlinearity and all values of the sweep rate with $v_0/\Delta^2 >0.1$, including the fast-passage regime. One issue that arises when we fix $v_1/(v_0\Delta)$ is that the condition $\chi_2\ll 1$ will necessarily be violated in the adiabatic limit. Therefore the perturbative calculation does not apply in that limit. More specifically, since we fix the ratio $v_1/(v_0\Delta)$ for each data set, the validity condition becomes $v_0/\Delta^2 \gg v_1/(v_0\Delta)$. For both $v_1/(v_0\Delta)=0.05$ and $v_1/(v_0\Delta)=0.02$, the theoretical formula clearly breaks down when $v_0/\Delta^2\lesssim v_1/(v_0\Delta)$. Taken together, Figs.~\ref{Fig:ProbabilityDeviationFixedNonlinearityWeak} and \ref{Fig:ProbabilityDeviationVariableNonlinearityWeak} show that if the quadratic term is small compared to the linear term throughout the crossing duration, the perturbative DDP formula provides a good approximation for the transition probability.

In addition to plotting the perturbative formula derived in Sec.~\ref{Sec:PerturbativeDDPQuadratic} for the lowest-order correction, we performed a numerical calculation of the DDP formula using the same sweep function that we used in the Schr\"odinger equation, i.e.~Eq.~(\ref{Eq:VariableRateSweepFunction}). In other words, solving the equation $\mathcal{E}(t)=0$ and evaluating the integral $D(t_c)$ were performed numerically using Eq.~(\ref{Eq:VariableRateSweepFunction}). The results of this calculation agree even better with the results of the numerical simulation of the dynamics (compare the dashed brown line with the red + symbols in Fig.~5). In particular, we no longer obtain the incorrect increase in $\delta P$ at small values of $v_0/\Delta^2$.

It is worth pausing at this point to consider the following note on the results in the perturbative regime: the analytic formula and numerical simulations were obtained using two different functions that are drastically different away from the crossing point. The fact that the results of the two approaches agree with each other means that the behavior away from the crossing point does not affect the final results even though one might expect the difference between the two functions to affect the zeros, defined by Eq.~(\ref{EquationForZeroPoints}), in the complex plane. After all, the zeros in the complex plane are obtained via analytic continuation, which depends on the function on the entire real line. It turns out, however, that the zeros and integrals in the DDP formula are most sensitive to $\epsilon(t)$ around $t=0$, and the two sweep functions are almost equal for small values of $|t|$.

In Fig.~\ref{Fig:ProbabilityDeviationStrong} we plot simulation results for stronger nonlinearity. The perturbative DDP formula [Eq.~(\ref{Eq:LZFormulaQuadratic})] is a good approximation for large values of $v_0/\Delta^2$, but it fails away from that regime. It is interesting that Eq.~(\ref{Eq:CorrectionToLZFormula}), which was derived as a simple approximation to Eq.~(\ref{Eq:LZFormulaQuadratic}), exhibits better agreement with the simulation results for the values of $v_0/\Delta^2$ that correspond to the largest deviations from $P_{\rm LZSM}$.

In Fig.~\ref{Fig:ProbabilityDeviationVeryStrong} we plot simulation results for an extremely strong nonlinearity along with results from a numerical calculations following the DDP approach. Remarkably the DDP formula still works but only after we include a sufficient number of zeros in the calculation. In particular, keeping only one zero of $\mathcal{E}(t)$ gives accurate results in the fast-passage limit but fails as we move towards the adiabatic limit. This feature can be understood by noting that a second zero becomes increasingly important as we approach the adiabatic limit. Keeping two zeros gives accurate results in the adiabatic limit. In fact, all the zeros of $\mathcal{E}(t)$ form pairs with asymptotically vanishing intra-pair distance in the adiabatic limit, which leads to the result that keeping an even number of zeros leads to better approximations in this limit. Apart from the odd-even difference, the approximations generally improve as we increase the number of zeros that we keep in the DDP calculation. If we keep five zeros in the calculation, we obtain good agreement with the Schr\"odinger equation solution everywhere. A point worth noting about Fig.~\ref{Fig:ProbabilityDeviationVeryStrong} is that when we keep two zeros we obtain $P>1$ in the fast-passage regime. This result is clearly unphysical. However, this seemingly serious problem is specific to the choice of an inappropriate number of zeros that we keep in the DDP calculation and does not imply that the DDP approach as a whole is invalid in this case. No similar behavior (i.e.~$P>1$) occurs for any other number of zeros.

Figure \ref{Fig:ProbabilityDeviationCubic} shows plots for the case of cubic nonlinearity. As with the case of a quadratic nonlinearity, when we fix the nonlinearity coefficient $\chi_3$, we obtain good agreement between the perturbative DDP formula and the exact results in the adiabatic regime, but the approximate formula breaks down in the fast-passage regime. When we use parameters such that $\chi_3$ decreases with increasing $v_0/\Delta^2$, we obtain good agreement between the approximate DDP formula and the exact results.

\subsection{Double-passage problem}
\label{Sec:DoublePassage}

We now go back to the step of locating the zeros of $\mathcal{E}(t)$ in the case of a quadratic nonlinearity. In Sec.~\ref{Sec:PerturbativeDDPQuadratic} we proceeded by determining the new, shifted location of the zero that exists in the linear case. However, if we inspect $\mathcal{E}(t)$ with the quadratic sweep function
\begin{equation}
\epsilon(t) = v_0 t + v_1 t^2 = -\frac{v_0^2}{4v_1} + v_1 \left( t + \frac{v_0}{2v_1} \right)^2
\label{Eq:ParabolicSweepFunction}
\end{equation}
more closely, we find that in addition to $t'_c$ there are three other zeros: $\left. t'_c \right.^*$, $(-v_0/v_1-t'_c)$ and $(-v_0/v_1-t'_c)^*$. (The asterisk denotes complex conjugation.) The last one of these has a positive imaginary part. We can then use the generalized DDP formula and include contributions from the zeros at $t'_c$ and $-v_0/v_1-\left. t'_c \right.^*$. We have already derived an approximate expression for $D(t'_c)$ in Sec.~\ref{Sec:PerturbativeDDPQuadratic}. By examining the integration path that goes from $s=0$ to $s=-v_0/v_1$ and then to $s=-v_0/v_1-\left. t'_c \right.^*$, we find that
\begin{equation}
D\left( -\frac{v_0}{v_1} - \left. t'_c\right.^* \right) = -D(t'_c)^* + \int_0^{-v_0/v_1} \mathcal{E}(s) ds.
\label{Eq:DDPDoublePassageRelation}
\end{equation}
The last integral can be recognized as the dynamical phase that accumulates between the two crossings in the double-passage problem. The imaginary parts of $D\left( -v_0/v_1 - \left. t'_c\right.^* \right)$ and $D(t'_c)$ are equal because the sweep function is symmetric: both the sweep rate and nonlinearity are equal at the two crossing points. The minus sign and complex conjugation in the first term on the right-hand side of Eq.~(\ref{Eq:DDPDoublePassageRelation}) mean that (excluding the dynamical-phase term) the real part of the $D\left( -v_0/v_1 - \left. t'_c\right.^* \right)$ has the opposite sign to the real part of $D(t'_c)$. The difference between the real parts of $D(t'_c)$ and $D\left( -v_0/v_1 - \left. t'_c\right.^* \right)$ can then be interpreted as a geometric phase accumulated between the two crossings. So far, these results seem to be consistent with our knowledge about the double-passage LZSM problem. However, when we substitute the relevant expressions in Eq.~(\ref{Eq:GeneralizedDDP}), we obtain the net transition probability
\begin{equation}
P \approx P_{\rm LZSM} \left(1 + \frac{3}{4}\pi\delta \chi_2^2 \right) \times \left| 1 - e^{i\phi} \right|^2,
\label{Eq:DoublePasssageProbabilityDDP}
\end{equation}
where
\begin{equation}
\phi = \frac{2}{3}\frac{\Delta^2}{v_0}\chi_2 - \int_0^{-v_0/v_1} \mathcal{E}(s) ds.
\end{equation}
This result is clearly problematic. Excluding the factor $|1-e^{i\phi}|^2$, Eq.~(\ref{Eq:DoublePasssageProbabilityDDP}) describes a function that increases from zero in the adiabatic limit to one in the fast-passage limit. The factor $|1-e^{i\phi}|$ lies between 0 and 2, depending on the phase $\phi$. Therefore, if we consider a double-passage problem with both passages being in the fast limit and a phase $\phi$ that is an odd-integer-multiple of $\pi$, we would obtain a probability that is approximately equal to 4, a clearly unphysical result. Meanwhile, in a symmetric double-passage problem, the transition probability should approach zero in both the adiabatic and fast-passage limits. It should be noted that the sweep function in Eq.~(\ref{Eq:ParabolicSweepFunction}) does not violate any of the conditions that we mentioned above for the validity of the DDP approach. It should also be noted that the solution to this paradoxical situation cannot be in including more zeros in the generalized DDP formula, because there are no additional zeros apart from the two that are included in Eq.~(\ref{Eq:DoublePasssageProbabilityDDP}). We do not have a definitive explanation for why the DDP approach fails in this case. We just note that numerical calculations suggest that the DDP approach gives good results in the regime $\delta\gtrsim 1$, which is the regime where the DDP calculation has rigorous justification.

It is also interesting that by keeping one zero we obtained correct results for a single passage with nonlinearity, although this approximation was not rigorously founded, while keeping both zeros did not produce good results for the double-passage problem, even though this procedure seems to be more consistent with the conventional wisdom in applying the DDP approach.

\section{Other related problems}
\label{Sec:OtherCases}

\subsection{Some examples of perturbatively nonlinear sweep functions}

We now consider a few special cases involving functions that allow us to make some progress with analytical derivations. First we treat the case of a sinusoidal function
\begin{equation}\label{sine}
\epsilon_1(t)=A\sin(t/T)
\end{equation}
together with a constant $\Delta$. If the time variable extends over a sufficiently long duration, the sine function is an oscillating function, which means that we will be dealing with a periodic driving problem \cite{ShevchenkoReview,Ashhab,Li}. The dynamics therefore exhibits oscillations that continue indefinitely, unlike the single-passage problem where the diabatic basis state probabilities approach constant asymptotic values at $t\rightarrow\infty$. As a result, this case does not fit the picture of a single-passage LZSM problem discussed in this manuscript. Nevertheless, as we shall see below, it is interesting to apply the DDP formula to this case and see what results this calculation gives.

At the crossing points, e.g.~taking $t=0$ for definiteness, the first three derivatives of $\epsilon_1(t)$ are given by $\dot{\epsilon}_1(0)=A/T$, $\ddot{\epsilon}_1(0)=0$ and $\epsilon_1^{(3)}(0)=-A/T^3$. Considering that $\sin(x+iy)=\sin x \cosh y + i \cos x \sinh y$, we find that the solutions of the equation
\begin{equation}
A^2 \sin^2\frac{t}{T} + \Delta^2 = 0
\end{equation}
are given by
\begin{equation}\label{tc}
t_c^{(n,\pm)}=n\pi T \pm i\nu,
\end{equation}
where $n$ is any integer,
\begin{equation}
\nu = T \, {\rm arcsinh} \, \xi = T\ln(\xi+\sqrt{\xi^2+1}),
\end{equation}
and the parameter 
\begin{equation}\label{xi}
\xi=\Delta/A
\end{equation}
quantifies the nonlinearity. Note that the parameter $\xi$ for the sweep function (\ref{sine}) is related to the parameter $\chi_3$ for the sweep function (\ref{FirstThird}) by the formula $\chi_3=-\xi^2$. As a first step, we make the arbitrary choice $n=0$, i.e. we choose $t_c^\prime = t_c^{(0,+)}=i\nu$. The integral
\begin{equation}\label{integral-1}
D(t_c^\prime)=\int_0^{i\nu}\sqrt{\Delta^2+A^2\sin^2(s/T)}\,ds,
\end{equation}
after the substitution $s=iTz$, reads
\begin{equation}\label{integral-2}
D(t_c^\prime)=iT\Delta\int_0^{\nu/T}\sqrt{1-\xi^{-2}\sinh^2 z}\,dz.
\end{equation}
Note that the integrand in Eq.~(\ref{integral-2}) decreases from 1 to 0 and is real throughout the integration interval. As a result, $D(t_c^\prime)$ is purely imaginary. The substitution $w = \sinh z$ gives
\begin{equation}\label{integral-3}
D(t_c^\prime)=iT\Delta\int_0^{\sinh(\nu/T)}\sqrt{\frac{1-\xi^{-2}w^2}{1+w^2}}\,dw.
\end{equation}
Taking into account that $\sinh(\nu/T)=\xi$ and making the substitution $u = w/\xi$, we obtain
\begin{eqnarray}\label{integral-4}
	D(t_c^\prime)&=&iT\Delta\;\xi\int_0^1\sqrt{\frac{1-u^2}{1+(\xi u)^2}}\,du\\
	&=&iT\Delta\frac{\sqrt{1+\xi^2}}{\xi}\nonumber\\
	&\times&\left[K\left(\frac{\xi}{\sqrt{1+\xi^2}}\right)-
	E\left(\frac{\xi}{\sqrt{1+\xi^2}}\right)\right],\nonumber
\end{eqnarray}
where $K(k)$ and $E(k)$ are the full elliptic integrals of the first and second kind, respectively.
For $\xi\ll 1$, the integral in Eq.~(\ref{integral-4}) can be approximated by expanding $1/\sqrt{1+(\xi u)^2}$:
\begin{equation}\label{integral-5}
D(t_c^\prime)=iT\Delta\;\xi\left\{\int_0^1\sqrt{1-u^2}\,du - \frac{1}{2}\xi^2\int_0^1 u^2 \sqrt{1-u^2}\,du\right\}.
\end{equation}
As
\begin{equation}\label{integral-6}
\int_0^1\sqrt{1-u^2}\,du=\frac{\pi}{4}, \qquad  \int_0^1 u^2 \sqrt{1-u^2}\,du=\frac{\pi}{16},
\end{equation}
the transition probability is given by Eq.~(\ref{Eq:LZFormulaCubic}). In other words, keeping only the linear and cubic terms in the sweep function gives the same lowest-order effect of the nonlinearity as working with the sine function without truncating its Taylor series expansion.

\begin{figure}[h]
\includegraphics[width=8.5cm]{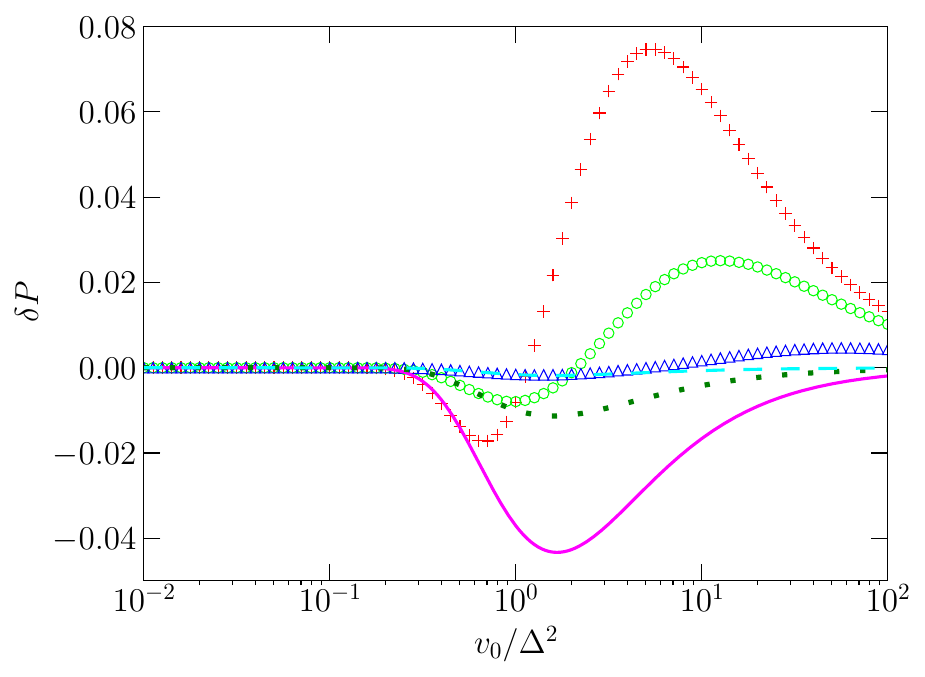}
\caption{Probability difference $\delta P$ as a function of $v_0/\Delta^2$ for the sweep function $\epsilon(t)=A\sinh(t/T)$. For this sweep function, the sweep rate at the crossing point is given by $v_0=A/T$. The red + symbols, green circles and blue triangles correspond, respectively, to $A/\Delta=1, 2$ and 5. The parameter $T$ is set at $T=A/v_0$ and is therefore not fixed for any of the data sets in the figure. The solid magenta, dotted green, and dashed cyan lines are obtained using Eq.~(\ref{Eq:LZFormulaCubic}) for $A/\Delta=1, 2$ and 5, respectively. Similar to the quadratic and cubic cases, the approximate formula agrees well with the exact results in the adiabatic regime, but the exact results exhibit peaks in the fast-passage regime that are not reproduced by the perturbative DDP formula.}
\label{Fig:ProbabilityDeviationSinhFixedNonlinearity}
\end{figure}

\begin{figure}[h]
\includegraphics[width=8.5cm]{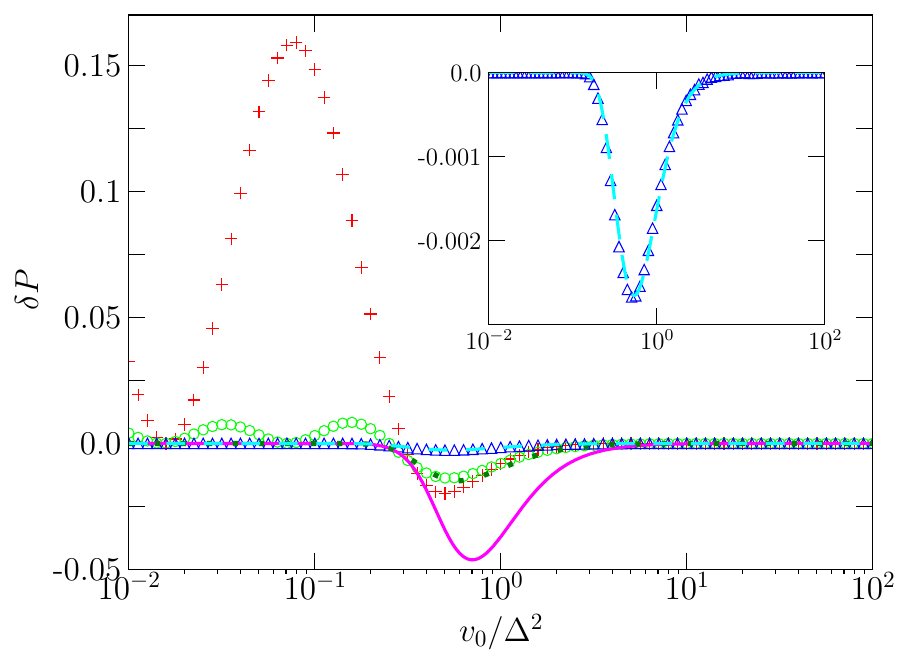}
\caption{Same as in Fig.~\ref{Fig:ProbabilityDeviationSinhFixedNonlinearity}, but now we use a variable $A$. The red + symbols, green circles and blue triangles correspond, respectively, to $A\Delta/v_0=1, 2$ and 5. The solid magenta, dotted green, and dashed cyan lines are obtained using Eq.~(\ref{Eq:LZFormulaCubic}) for $A\Delta/v_0=1, 2$ and 5. The weakest nonlinearity case ($A\Delta/v_0=5$; shown also in the inset) exhibits good agreement between the perturbative DDP formula and the exact results.}
\label{Fig:ProbabilityDeviationSinhVariableNonlinearity}
\end{figure}

Analogous calculations can be performed for 
\begin{equation}
\epsilon_2(t)=A\sinh(t/T)
\end{equation} 
(see Appendix B) and 
\begin{equation}
\epsilon_3(t)=A\tanh(t/T).
\end{equation} 
We again find that in the perturbative regime ($\xi\ll 1$) the correction is given by the formula in Eq.~(\ref{Eq:LZFormulaCubic}). In accordance with the results of Ref.~\cite{Vitanov}, $P<P_{\rm LZSM}$ for the superlinear driving function $\epsilon_2(t)$ and $P>P_{\rm LZSM}$ for the sublinear driving functions $\epsilon_1(t)$ and $\epsilon_3(t)$. Moreover, our result for $\epsilon_2(t)$ coincides with Eq.~(19) of Ref.~\cite{Vitanov} obtained for the superlinear function in Eq.~(\ref{superlinear}), and also our results for $\epsilon_1(t)$ and $\epsilon_3(t)$ coincide with Eq.~(29a) of Ref.~\cite{Vitanov} obtained for the sublinear function in Eq.~(\ref{sublinear}).

In Figs.~\ref{Fig:ProbabilityDeviationSinhFixedNonlinearity} and \ref{Fig:ProbabilityDeviationSinhVariableNonlinearity}, we plot $\delta P$ for the case of $\epsilon_2(t)=A\sinh(t/T)$. In Fig.~\ref{Fig:ProbabilityDeviationSinhFixedNonlinearity}, we keep the nonlinearity coefficient $\chi_3$ fixed for each data set. In other words, for each value of $A$, the parameter $T$ is given by $T=A/v_0$. As with the case of a linear-plus-cubic sweep function, we obtain peaks in $\delta P$ for the exact results in the fast-passage regime, signaling a breakdown of the perturbative DDP calculation. If we avoid this breakdown by setting $A\Delta/v_0$ to a fixed value for each data set, i.e.~setting $A$ to be proportional to $v_0/\Delta^2$ (and keeping the relation $T=A/v_0$), we obtain good agreement between the approximate formula and exact results everywhere for a sufficiently weak nonlinearity [Fig.~\ref{Fig:ProbabilityDeviationSinhVariableNonlinearity}]. For stronger nonlinearity, the agreement remains rather poor, especially in the adiabatic regime, where now $A/\Delta$ becomes small and violates the condition of weak nonlinearity.

\subsection{Uniformly rotating field}

\begin{figure}[h]
\includegraphics[width=8.5cm]{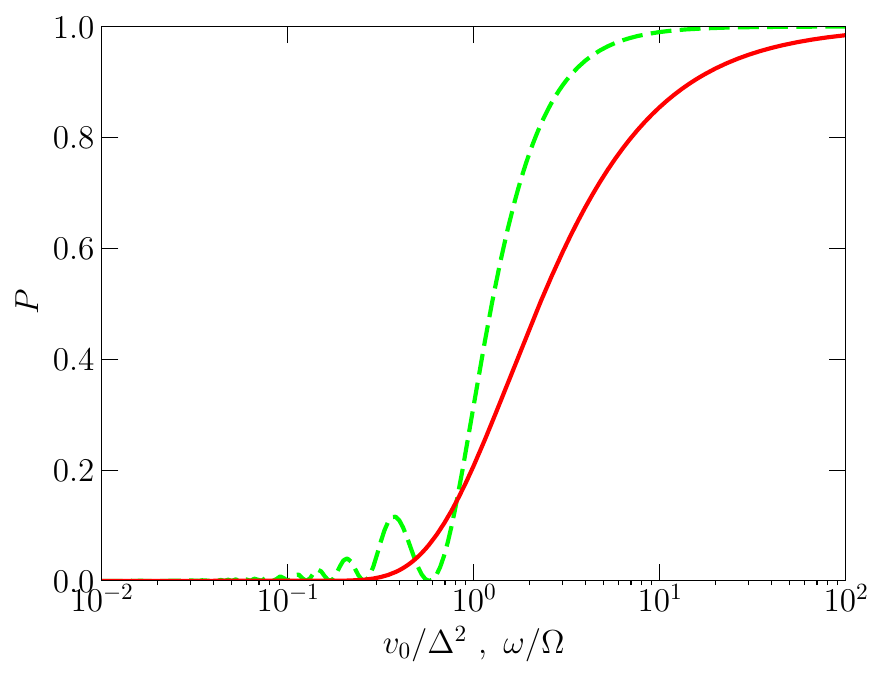}
\caption{Transition probability $P$ for a uniformly rotating field of fixed magnitude. The dashed green line shows the function in Eq.~(\ref{Eq:RotatingFieldSingleFlip}). The solid red line shows $P_{\rm LZSM}$ for reference. To compare these two cases we make the correspondence $\Delta \leftrightarrow \Omega$ and $v_0 \leftrightarrow \omega \Omega$, which give $v_0/\Delta^2 \leftrightarrow \omega / \Omega$.}
\label{Fig:ProbabilityRotatingField}
\end{figure}

We now consider the problem of a two-level system in an external field that has a fixed magnitude and rotates at constant speed:
\begin{eqnarray}\label{RotatingFieldFunctions}
	\epsilon(t) & = & \Omega \cos \omega t, \nonumber \\
	\Delta(t) & = & \Omega \sin \omega t,
\end{eqnarray}
from the initial time $t=0$ until a final time $t=T$. The problem can be transformed into one with a fixed external field by making a reference frame transformation to a frame that rotates about the $y$ axis with frequency $\omega$, such that the external field always points along the $z$ axis. This transformation is similar to the standard one used in the study of Rabi oscillations. The dynamics is then governed by the Schr\"odinger equation with the effective Hamiltonian:
\begin{equation}
	\tilde{H} = \frac{1}{2} \left( \begin{array}{cc} \Omega & i \omega \\ -i \omega & - \Omega \end{array} \right).
\end{equation}
Straightforward algebra then shows that for an initial state that is an eigenstate of the initial Hamiltonian, e.g.~the ground state, the probability for the system to make a transition and end up in the other adiabatic state at a later time $t$, i.e.~the excited state of the Hamiltonian $H(t)$, is given by:
\begin{equation}
	P = \frac{1}{2} \frac{x^2}{1+x^2} \left[ 1 - \cos \left( \sqrt{\Omega^2 + \omega^2} t \right) \right],
	\label{Eq:RotatingFieldProbability}
\end{equation}
where the parameter $x=\omega/\Omega$ quantifies the adiabaticity of the Hamiltonian variation. As one would intuitively expect, the adiabatic limit $x\rightarrow 0$ gives $P=0$. The fast-rotation limit $x\rightarrow \infty$ (i.e.~$\omega\gg\Omega$) gives $P=[1-\cos(\omega t)]/2$, which oscillates between 0 and 1 at the same frequency as the rotating field. This result can be understood as the quantum system being unable to react to the fast-oscillating field and hence remaining frozen in the initial state in the lab frame, which also agrees with the intuitive expectation. If we take the special case in which the field makes a single 180\textdegree{\,} rotation from the beginning to the end of the field variation, the transition probability is given by Eq.~(\ref{Eq:RotatingFieldProbability}) with $t=\pi/\omega$:
\begin{equation}
	P = \frac{1}{2} \frac{x^2}{1+x^2} \left[ 1 - \cos \left( \sqrt{1+x^{-2}} \pi \right) \right],
	\label{Eq:RotatingFieldSingleFlip}
\end{equation}
which exhibits small oscillations at intermediate values of $x$ but is well behaved in the limit $x\rightarrow \infty$ and asymptotically approaches 1 (see Fig.~\ref{Fig:ProbabilityRotatingField}).

If we try to apply the DDP formula to this problem, we obtain the equation
\begin{equation}
	\mathcal{E}(t) = \Omega \sqrt{\cos^2 \omega t + \sin^2 \omega t} = 0,
\end{equation}
which clearly does not have any solutions. Hence, the DDP formula cannot be used, even though this problem is closely related to LZSM problems. It is not entirely surprising that the DDP formula does not work here. If we consider the time variable extending from $-\infty$ to $\infty$, with $\epsilon$ and $\Delta$ being fixed before $t=0$ and after $t=\pi/\omega$, then the functions $\epsilon(t)$ and $\Delta(t)$ are non-analytic, since they start and stop oscillating abruptly.

\section{Essential nonlinearity}
\label{Sec:EssentialNonlinearity}

\begin{figure}[ht!]
\includegraphics[width=8.5cm]{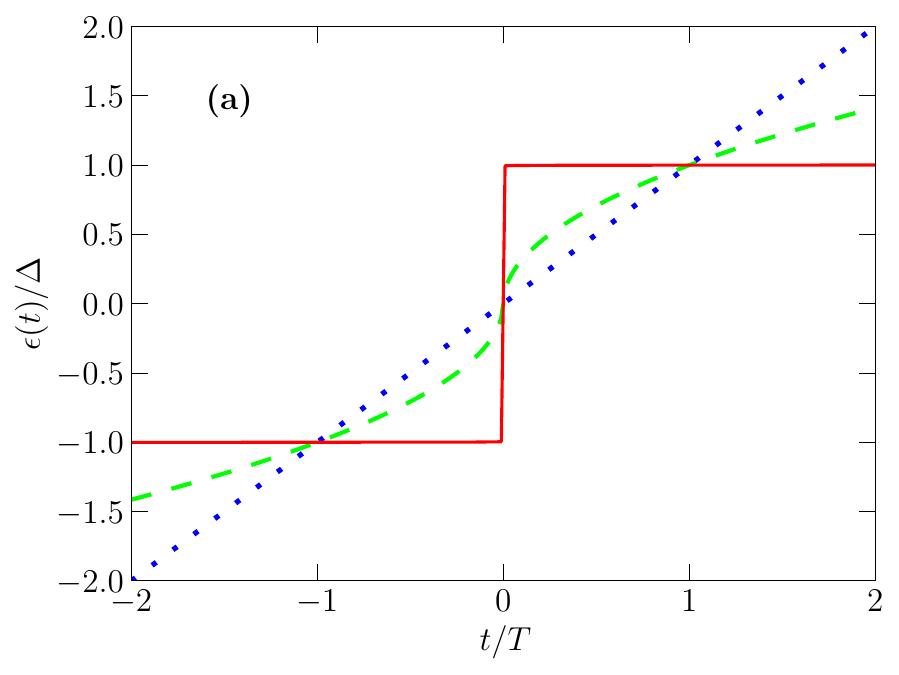}
\includegraphics[width=8.5cm]{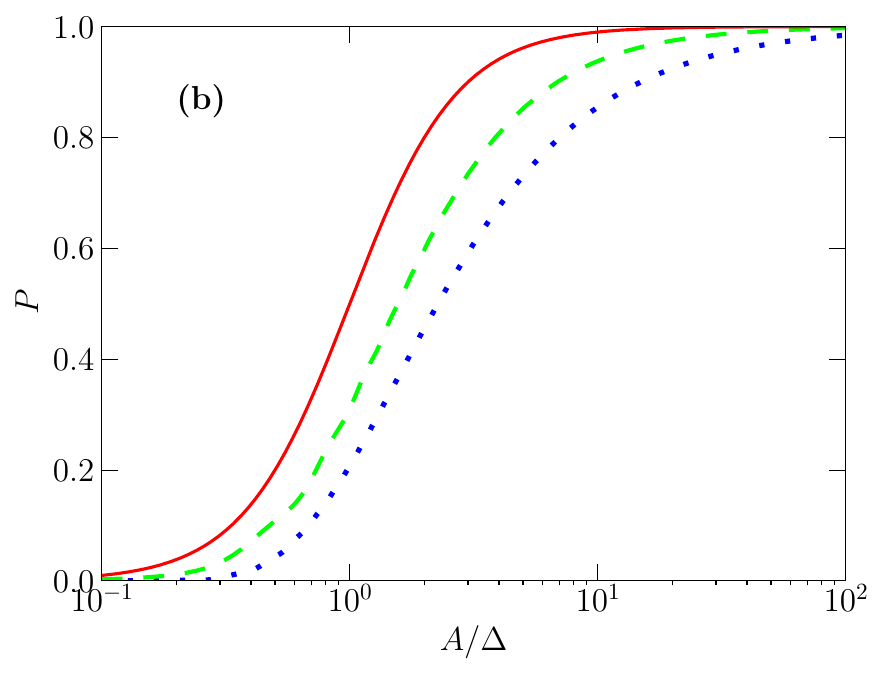}
\caption{(a) The sweep function $\epsilon(t)$, Eq.~(\ref{power}), as a function of time $t$ for $A/\Delta=1$ and $a=10^{-3}$ (solid red line), 0.5 (dashed green line), 1 (dotted blue line). (b) The numerically calculated transition probability as a function of the amplitude $A$ (measured relative to $\Delta$). The dotted blue and dashed green lines correspond, respectively, to $a=1$ and 0.5 in Eq.~(\ref{power}). The solid red line shows three virtually coinciding (indistinguishable) curves --- for $a=10^{-3}$ in Eq.~(\ref{power}), for $\sigma=10^{-3}$ in Eq.~(\ref{error-function}), and for the limiting function (\ref{limiting-function}).}
\label{Fig:Essential_Nonlinearity}
\end{figure}

The sweep function
\begin{equation}\label{power}
\epsilon(t)=A\,\text{sgn}(t) \left|\frac{t}{T}\right|^a, \qquad a>0
\end{equation}
is essentially nonlinear for $a\neq 1$: $\dot{\epsilon}(0)=0$ for $a>1$ and 
$\dot{\epsilon}(0)=\infty$ for $0<a<1$. For non-integer $a$, the function (\ref{power}) is nonanalytical, and the generalized DDP formula cannot be used. We therefore calculate the LZSM probability numerically by solving the Schr\"odinger equation. In the present study, we are especially interested in $a\ll 1$, when Eq.~(\ref{power}) describes an almost square pulse. Periodic square pulses are studied in literature, both theoretically and experimentally (see, for example, Refs.~\cite{SilveriReview,Silveri2015}). In Fig.~\ref{Fig:Essential_Nonlinearity}, we show the sweep functions for several values of $a$ and the dependence of the probability $P$, calculated numerically, on $A/\Delta$. For $a\to 0$, the numerically calculated occupation probability $P$ as a function of the amplitude $A$ [see Fig.~\ref{Fig:Essential_Nonlinearity}(b)] tends to the limiting function
\begin{equation}\label{limiting-function}
P_{\text{lim}}(A)=\frac{A^2}{\Delta^2+A^2}.
\end{equation}
Interestingly, the error function
\begin{equation}\label{error-function}
\epsilon(t)=A\,\text{erf}\left(\frac{t}{\sqrt{2}\sigma T}\right)=
A\int_{0}^{t/\sqrt{2}\sigma T}e^{-\tau^2}d\tau,
\end{equation}
which has a finite derivative $\dot{\epsilon}(0)=A/\sqrt{2}\sigma T$, also describes a square pulse for $\sigma\ll 1$, and, for $\sigma\to 0$, $P_{\text{erf}}(\sigma)$ tends to the same limiting function (\ref{limiting-function}). We compare the occupation probability $P$ for $a=10^{-3}$, 0.5, 1 and $P_{\text{erf}}$ for $\sigma=10^{-3}$ in one plot in Fig.~\ref{Fig:Essential_Nonlinearity}(b).

\section{Time-dependent gap}
\label{Sec:VariableDelta}

\subsection{Eliminating the time dependence of $\Delta(t)$}

We now consider the qubit Hamiltonian 
\begin{equation}
	H = \frac{1}{2} \left( \begin{array}{cc} \epsilon(t) & \Delta(t) \\ \Delta(t) & - \epsilon(t) \end{array} \right),
	\label{Eq:LZHamiltonian}
\end{equation}
with a general sweep function $\epsilon(t)$ and a time-dependent gap $\Delta(t)$.

With an appropriate transformation of the time variable, the time-dependent Schr\"odinger equation can be transformed to a form that has a time-independent gap and a modified sweep function. The details of the derivation are shown in Appendix~\ref{App:TimeDependentGap}.

Once the problem is transformed to a time-independent-gap LZSM problem, the formulae that we derived in previous sections can be applied. In particular, if we consider the case of a linear sweep function $\epsilon(t)=vt$ and a weakly time-dependent gap [$\Delta(t)=\Delta_0+\Delta't$ in the vicinity of the crossing point], we obtain the perturbative formula
\begin{equation}
P \approx \exp \left\{-2\pi\delta \left(1 - \frac{3(\Delta')^2}{2v^2} \right) \right\}.
\label{Eq:LZFormulaVariableDelta}
\end{equation}
The line in Fig.~\ref{Fig:ProbabilityDeviationVariableNonlinearityWeak} that corresponds to a linear sweep function and time-dependent $\Delta$ is fitted well by this formula.

\subsection{Asymptotically vanishing gap}
\label{Sec:VanishingGap}

In this section, we consider a different variation on the LZSM problem that can arise naturally in realistic systems, namely the situation where $\Delta$ has a maximum at the avoided crossing point and decreases to much smaller values when $\epsilon\rightarrow\pm\infty$. For example, if one thinks of a problem described in terms of a single particle trapped in a time-dependent potential well, the trapping potential can be deformed in time such that two things occur simultaneously: (1) the energies of local minima in two local wells move up and down relative to each other such that the locations of the ground and first-excited states switch, and (2) the distance and/or energy barrier between the two wells increase away from the crossing point, such that the energy scale coupling the states in the two wells decreases away from the crossing point. This situation can occur, for example, in the context of superconducting qubits, where the effective potential for the phase variables varies in complex ways if any of the bias parameters is varied.

\begin{figure}[h]
	\includegraphics[width=8.5cm]{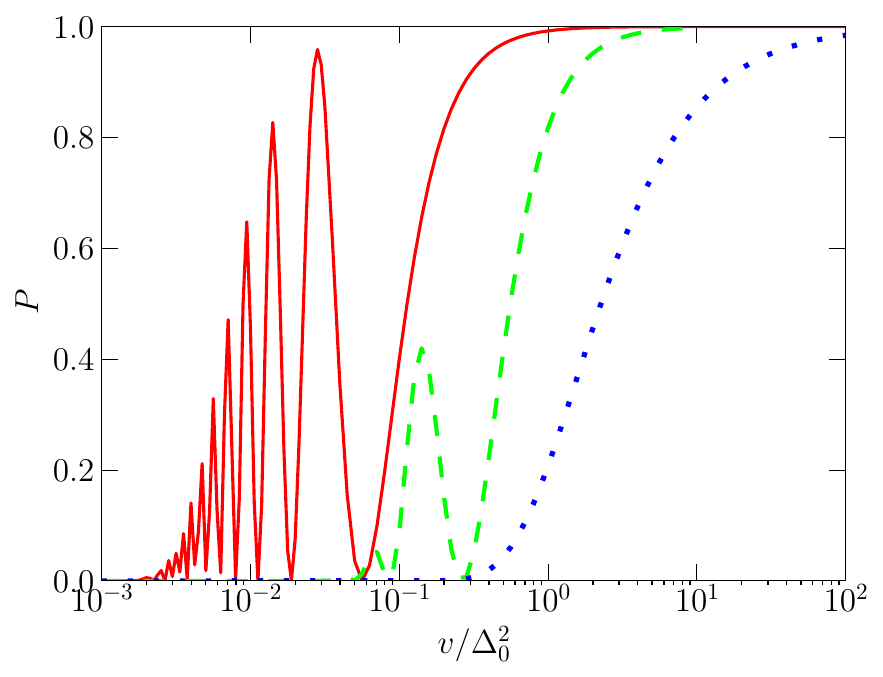}
	\caption{Transition probability in the case of an asymptotically vanishing gap, described by Eq.~(\ref{Eq:GaussianGap}). The solid red line corresponds to $T=0.1/\Delta_0$. The dashed green line corresponds to $T=0.5/\Delta_0$. The dotted blue line corresponds to the original LZSM problem, which also corresponds to the limit $T\rightarrow\infty$. The red line in this figure does not go all the way down to zero in each oscillation because of the limited number of points in our data set.}
	\label{Fig:ProbabilityVariableGap}
\end{figure}

For definiteness, we consider a Gaussian function when describing the suppression of $\Delta$ away from the crossing point, i.e.~
\begin{eqnarray}
	\epsilon(t) & = & v t,
	\nonumber \\
	\Delta(t) & = & \Delta_0 e^{-(t/T)^2}.
	\label{Eq:GaussianGap}
\end{eqnarray}
The results of numerical simulations are shown in Fig.~\ref{Fig:ProbabilityVariableGap}. We also performed calculations using a Lorentzian function, and these calculations produced qualitatively similar results.

When $\Delta(t)$ decreases quickly away from the crossing point, the transition probability increases in general. This effect resembles the effect of having a smaller value of $\Delta$. This result is logical, since having $\Delta$ active for a shorter duration can be expected to result in a smaller effective value of $\Delta$. The transition probability also exhibits more oscillatory behavior with decreasing width $T$. This result is also to be expected, since the presence of oscillations is natural for pulsed manipulation of quantum systems, while the absence of oscillations in the linear LZSM problem is a rather particular feature of that problem.

\section{Conclusions}
\label{Sec:Conclusion}

Realizations of the energy-level avoided crossings are ubiquitous in quantum physics, and the driving of quantum systems through such energy level structures is usually described by the linearized model, where a generally nonlinear function $\epsilon(t)$ is replaced by a linear one, $\epsilon(t)=vt$. But what is the effect of this approximation, and what is neglected by the linearization procedure? We have addressed this question, which is becoming increasingly important in recent years, given the precision of modern quantum technologies. One of the important results that we have demonstrated in this manuscript is that the corrections to the LZSM formula under the influence of a perturbatively nonlinear parameter variation are smaller than what one might intuitively expect. Besides demonstrating analytically and numerically the robustness of the LZSM formula, our corrections may become important for realistic systems. Our results demonstrate that the choice of a nonlinear driving function and tuning its parameters can be used for alternative quantum control protocols.

The main results can be summarized as follows:

(1) We have obtained analytically, using the DDP formula, the first correction to the LZSM probability for several perturbatively nonlinear sweep functions: linear-plus-quadratic, linear-plus-qubic, sine, sinh and tanh functions in the single-passage regime. We have shown that the correction for the odd sine, sinh and tanh functions is given by the same formula, Eq.~(\ref{Eq:LZFormulaCubic}), as for the linear-plus-qubic function based on the Taylor expansion.

(2) We have compared the perturbative analytical calculations with results obtained by numerical solving of the Schr\"odinger equation. The agreement is very good for weak nonlinearity, i.e.~for parameters for which the nonlinear term in the sweep function is small compared to the linear term in the crossing region. The crossing region here corresponds to the time interval during which the state probabilities experience significant changes, as established in previous studies on LZSM dynamics. The approximate expressions become invalid if this condition of small nonlinearity is violated.

(3) We have analyzed the double-passage problem for the linear-plus-quadratic sweep function and found that the DDP approach is not applicable in this case.

(4) We have considered the case of the sweep function $\epsilon(t)=v_0 t (1+\alpha \tanh t/T)$ to study a case of strong nonlinearity. The numerical evaluation of the transition probability following the DDP approach gives good agreement with the results obtained by numerically solving the Schr\"odinger equation when a sufficiently large number of zeros, Eq.~(\ref{EquationForZeroPoints}), in the complex plane are taken into account.

(5) We have obtained an analytical formula for the transition probability for the case of a uniformly rotating field when the sweep function and the gap are given by Eq.~(\ref{RotatingFieldFunctions}).

(6) We have calculated numerically the transition probability for two essentially nonlinear sweep functions in the limit when they describe an almost square pulse --- the power function (\ref{power}) for $a\to 0$ and the error-function (\ref{error-function}) for $\sigma\to 0$. We have also given the limiting function for the probability in these cases.

(7) We have proven that the time dependence of the gap in the qubit Hamiltonian can be eliminated with a transformation of the time variable, such that the LZSM problem is reduced to one with a time-independent gap.

(8) We have studied numerically the LZSM problem with asymptotically vanishing gap (\ref{Eq:GaussianGap}) and found oscillations in the transition probability. The number of peaks and their maximum values grow with decreasing gap pulse width.

\section*{Acknowledgments}

We would like to thank O.M.~Bahrova, M.~Fujiwara, O.V.~Ivakhnenko, F.~Yoshihara and K.~Semba for useful discussions. S.A. was supported by Japan's MEXT Quantum Leap Flagship Program Grant Number JPMXS0120319794. O.A.I. was supported by NATO Science for Peace (Project G5796). S.N.S. was supported by the United States Army Research Office (Grant Number W911NF-20-1-0261).

\begin{appendix}
\renewcommand{\thesection}{\Alph{section}}

\section{Calculating $D(t'_c)$ for the case of a weak quadratic nonlinearity}
\label{App:QuadraticNonlinearity}

To illustrate one technique that we use for evaluating $D(t'_c)$, we first revisit the calculation in the linear case. The calculation of $D(t_c)$ in Eq.~(\ref{Eq:DtcForLinearLZSM}) can alternatively be performed as follows
\begin{eqnarray}
D(t_c) & = & i\int_0^{\tau_c}\sqrt{\Delta^2-(vs)^2} ds \nonumber \\
& = & i\int_0^{\tau_c}\sqrt{\Delta^2-(v(\tau_c - s'))^2} ds' \nonumber \\
& = & i\int_0^{\tau_c}\sqrt{v^2 \left(2 \tau_c s' - s'^2 \right)} ds' \nonumber \\
& = & i v \tau_c^2 \int_0^{1}\sqrt{2 x - x^2} dx 
= i v \tau_c^2 \times \frac{\pi}{4},
\end{eqnarray}
where we have defined $\tau_c=\Delta/v$. Reversing the roles of the lower and upper limits of the integral simplifies the treatment of the perturbation term that we shall add shortly. In particular, with this change we avoid complications that can arise with the Taylor expansion of the square root at the point where the square root vanishes.

The weak quadratic nonlinearity term modifies $\mathcal{E}(t)$,
\begin{equation}
\mathcal{E}(t) = \sqrt{\left( v_0 t + v_1 t^2 \right)^2 + \Delta^2},
\end{equation}
and shifts the zero of the DDP calculation from $t_c=i\Delta/v_0$ to $v_0 t'_c + v_1 (t'_c)^2=i\Delta$, which gives
\begin{eqnarray}
t'_c & = & \frac{-v_0 + \sqrt{v_0^2 + 4 i v_1 \Delta}}{2v_1}
\nonumber \\
& \approx & i \frac{\Delta}{v_0} + \frac{\Delta^2 v_1}{v_0^3} - i \frac{2\Delta^3 v_1^2}{v_0^5}.
\end{eqnarray}
We shall refer to $-it'_c$ as $\tau'_c$ below.

The integral $D(t'_c)$ now becomes
\begin{widetext}
\begin{eqnarray}
D(t'_c) & = & \int_0^{t'_c} \sqrt{\Delta^2 + (v_0 s + v_1 s^2)^2} ds
\nonumber \\
& = & \int_0^{t'_c} \sqrt{\Delta^2 + (v_0 (t'_c-s') + v_1 (t'_c-s')^2)^2} ds'
\nonumber \\
& = & \int_0^{t'_c} \sqrt{- v_0^2 (2 t'_c s' - s'^2 ) - 2 v_0 v_1 (3 (t'_c)^2 s' - 3 t'_c s'^2 + s'^3) - v_1^2 (4 (t'_c)^3 s' - 6 (t'_c)^2 s'^2 + 4 t'_c s'^3 - s'^4)} ds'
\nonumber \\
& = & i\int_0^{\tau'_c} \sqrt{v_0^2 (2 \tau'_c s' - s'^2 ) + 2 i v_0 v_1 (3 (\tau'_c)^2 s' - 3 \tau'_c s'^2 + s'^3) - v_1^2 (4 (\tau'_c)^3 s' - 6 (\tau'_c)^2 s'^2 + 4 \tau'_c s'^3 - s'^4)} ds'
\nonumber \\
& \approx & i\int_0^{\tau'_c} \sqrt{v_0^2 (2 \tau'_c s' - s'^2 )} ds' + i\int_0^{\tau'_c} \frac{2 i v_0 v_1 (3 (\tau'_c)^2 s' - 3 \tau'_c s'^2 + s'^3) - v_1^2 (4 (\tau'_c)^3 s' - 6 (\tau'_c)^2 s'^2 + 4 \tau'_c s'^3 - s'^4)}{2 v_0 \sqrt{2 \tau'_c s' - s'^2}} ds'
\nonumber \\
& & - i\int_0^{\tau'_c} \frac{-4 v_0^2 v_1^2 (3 (\tau'_c)^2 s' - 3 \tau'_c s'^2 + s'^3)^2}{8 v_0^3 (2 \tau'_c s' - s'^2)^{3/2}} ds'
\nonumber \\
& = & i v_0 (\tau'_c)^2 \int_0^{1} \sqrt{2 x - x^2} dx - v_1 (\tau'_c)^3 \int_0^{1} \frac{3 x - 3 x^2 + x^3}{\sqrt{2 x - x^2}} dx - i \frac{v_1^2}{2 v_0} (\tau'_c)^4 \int_0^{1} \frac{4 x - 6 x^2 + 4 x^3 - x^4}{\sqrt{2 x - x^2}} dx
\nonumber \\
& & + i \frac{v_1^2}{2 v_0} (\tau'_c)^4 \int_0^{1} \frac{(3 x - 3 x^2 + x^3)^2}{(2 x - x^2)^{3/2}} dx
\nonumber \\
& \approx & \frac{i\pi v_0}{4} \left( \frac{\Delta^2}{v_0^2} - 2i \frac{\Delta^3 v_1}{v_0^4} - 4 \frac{\Delta^4 v_1^2}{v_0^6} - \frac{\Delta^4 v_1^2}{v_0^6} \right) - \frac{3\pi - 4}{6} v_1 \left( \frac{\Delta^3}{v_0^3} - 3i \frac{\Delta^4 v_1}{v_0^5} \right) - \frac{5\pi}{32} i \frac{\Delta^4 v_1^2}{v_0^5} + \left( 2 - \frac{15\pi}{32} \right) i \frac{\Delta^4 v_1^2}{v_0^5}
\nonumber \\
& \approx & i \left( \frac{\pi\Delta^2}{4v_0} - \frac{3\pi}{8} \frac{\Delta^4 v_1^2}{v_0^5} \right) + \frac{2\Delta^3 v_1}{3v_0^3}.
\end{eqnarray}
\end{widetext}
Note that the step where we approximated the square-root using the Taylor expansion is valid when the first term inside the square root is much larger than the second and third terms combined, which implies the relation $v_1\ll v_0^2/\Delta$ or in other words $v_1\Delta/v_0^2\ll 1$.

\section{Calculation of the LZSM probability for the sinh function}

We consider the sweep function
\begin{equation}
\epsilon_2(t)= A \sinh\left(\frac{t}{T}\right).	
\end{equation}
Considering the Taylor expansion of the $\sinh$ function, the sweep function $\epsilon_2(t)$ gives
\begin{equation}
\chi_3 = \frac{\Delta^2}{A^2}.
\end{equation}
We now proceed with the DDP calculation for this case. Considering that $\sinh(x+iy)=\sinh x \cos y + i \cosh x \sin y$, we find that we need to specify the value of $\xi=\Delta/A$ to proceed with finding the solutions of the equation
\begin{equation}\label{zeros-sinh}
	A^2 \sinh^2\frac{t}{T} + \Delta^2 = 0.
\end{equation}
If $\xi<1$, all the solutions of Eq.~(\ref{zeros-sinh}) are purely imaginary and can be described by the formula
\begin{equation}
	t_c^{(n,\pm)} = \pm i T \arcsin\xi + i \pi T n,
\end{equation}
with $n$ being any integer. If $\xi>1$, the solutions are given by
\begin{equation}
	t_c^{(n,\pm)}=\pm \nu + i T \frac{\pi}{2} (2n+1),
\end{equation}
with
\begin{equation}
	\nu = T \, {\rm arccosh} \, \xi = T\ln(\xi+\sqrt{\xi^2-1}).
\end{equation}
For $\xi=1$, the roots of Eq.~(\ref{zeros-sinh}) have multiplicity 2 and are located at
\begin{equation}
	t_c^{(n)}= i T \frac{\pi}{2} (2n+1).
\end{equation}
If $\xi$ approaches 1 from below, each pair of zeros close to $iT\pi(n+1/2)$ approach each other and converge to the same point. If $\xi$ increases above 1, each zero splits back into two zeros that move away from each other in the horizontal direction.

For $\xi<1$, the solution of Eq.~(\ref{zeros-sinh}) that is closest to the real axis and has a positive imaginary part reads 
\begin{equation}
t_c^\prime = i T \arcsin\xi.
\end{equation}
Straightforward calculations [as in Section~\ref{Sec:OtherCases} for the sine function~(\ref{sine})] lead to the formula [analogous to Eq.~(\ref{integral-3})]
\begin{equation}
	D(t_c^\prime)=iT\Delta\int_0^{\xi}\sqrt{\frac{1-\xi^{-2}w^2}{1-w^2}}\,dw,
\end{equation}
and finally, in the limiting case $\xi\ll 1$, we get Eq.~(\ref{Eq:LZFormulaCubic}) for the transition probability. Note that if we take the limit $A/\Delta\to\infty$ and use the relation $T=A/v_0$ we obtain the probability $P_{{\rm LZSM}}$, as expected. 

For $\xi>1$, there are two solutions of Eq.~(\ref{zeros-sinh}) with a positive imaginary part that are the nearest ones to the real axis
\begin{equation}
	t_c^\pm=\pm T \ln(\xi + \sqrt{\xi^2 - 1})+i\frac{\pi}{2}T.
\end{equation}
Using the generalized DDP formula~(\ref{Eq:GeneralizedDDP}) we obtain for the probability
\begin{equation}
	P=2\exp{\left[-2\,{\rm Im}D(t_c^+)\right]}
	\Big\{1+\cos\left[2\,{\rm Re}D(t_c^+)\right]\Big\},
\end{equation}
where 
\begin{eqnarray}\label{ReD}
	{\rm Re}D(t_c^+)&=& A T 
	\Big\{\int_0^{\ln(\xi + \sqrt{\xi^2 - 1})}
	\sqrt{\xi^2 + \sinh^2 x}dx \nonumber\\
	&-&
	\int_0^{\pi/2}r(y,\xi)dy\Big\},
\end{eqnarray}
\begin{equation}\label{ImD}
	{\rm Im}D(t_c^+)=A T \int_0^{\pi/2}j(y,\xi)dy,
\end{equation}
\begin{equation}
	j(y,\xi)=
	\sqrt{\frac{u_1(y,\xi)+\sqrt{u_1^2(y,\xi)+u_2^2(y,\xi)}}{2}},
\end{equation}
\begin{equation}
r(y,\xi)=\frac{u_2(y,\xi)}{2j(y,\xi)},
\end{equation}
and
\begin{eqnarray}
	u_1(y,\xi)&=&\frac{1}{2}\Big[(2\xi^2-1)\cos(2y)-1\Big], \\ 
	u_2(y,\xi)&=&\xi\sqrt{\xi^2-1}\sin(2y).
\end{eqnarray}
The integrals in Eqs.~(\ref{ReD}), (\ref{ImD}) can be evaluated numerically. 

As mentioned above, the roots of Eq.~(\ref{zeros-sinh}) have multiplicity 2 at $\xi=1$, which violates one of the conditions required for the DDP formula. This point is interesting, because it suggests that there might be something pathological about the case $\xi=1$. However, in reality no unusual behavior occurs at this point, as confirmed by numerical simulations based on solving the Schr\"odinger equation. This point therefore illustrates one of the limitations of the DDP approach.

\section{Time-dependent gap $\Delta(t)$}
\label{App:TimeDependentGap}

In this Appendix, we show the details for eliminating the time dependence of the gap $\Delta(t)$ and derive the perturbative formula in this case.

\subsection{Transformation for eliminating the time dependence of $\Delta(t)$}

Let us take the time-dependent Hamiltonian~(\ref{Eq:LZHamiltonian}). We seek a transformation that eliminates the time dependence in $\Delta(t)$, i.e.~a mapping between the present problem and one with a time-independent $\Delta$. Below we present two equivalent methods that use slightly different languages.

\subsubsection{Method 1}

Let us make a substitution
\begin{equation}\label{G}
	t = G(\tilde t)
\end{equation}
in the Schr\"odinger equation (\ref{Eq:TDSE}) with the Hamiltonian~(\ref{Eq:LZHamiltonian}) such that
\begin{equation}\label{const}
	\frac{dt}{d\tilde t}\Delta(t)=\tilde{\Delta}=const.
\end{equation}
Equation (\ref{const}) can alternatively be expressed as 
\begin{equation}\label{substitution}
	G'(\tilde t)\Delta(G(\tilde t))=\tilde{\Delta}.
\end{equation}
After multiplying Eq.~(\ref{Eq:TDSE}) by $dt/d\tilde t=G'(\tilde t)$ we obtain
\begin{equation}\label{SchroedingerNew}
	i\frac{d|\psi(G(\tilde t))\rangle}{d\tilde t}=\tilde{H}(\tilde t)|\psi(G(\tilde t))\rangle,
\end{equation}
where the new Hamiltonian reads 
\begin{equation}\label{Hnew}
	\tilde{H}(\tilde t)=\frac{\tilde{\Delta}}{2}\sigma_{x}+\frac{\tilde\epsilon (\tilde t)}{2}\sigma_{z},
\end{equation}
with 
\begin{equation}\label{epsilon-new}
	\tilde\epsilon (\tilde t)=\epsilon(G(\tilde t))G'(\tilde t).  
\end{equation}
We look for a function $G(\tilde t)$ that satisfies Eq.~(\ref{substitution}), which for $G(0)=0$, is equivalent to
\begin{equation}\label{IntegralForm}
	\int_0^{G(\tilde t)}\Delta(t)dt=\tilde{\Delta} \tilde t.
\end{equation}
Assuming that $\Delta(t)>0$ and $\int_0^\infty\Delta(t)dt=\infty$, we conclude that there exists a unique solution $G(\tilde t)$ of Eq.~(\ref{substitution}). Therefore, the problem is reduced to one with a time-independent gap $\tilde\Delta$ and a modified sweep function $\tilde\epsilon(\tilde t)$, Eq.~(\ref{epsilon-new}).

\subsubsection{Method 2}

We define a new time-like variable $\tilde{t}$. The time-dependent Schr\"odinger equation can then be expressed as
\begin{equation}
	i\frac{d\ket{\psi}}{d\tilde{t}} = \frac{dt}{d\tilde{t}} H \ket{\psi}.
	\label{Eq:IntermediateTDSE}
\end{equation}
If we choose the relationship between $t$ and $\tilde{t}$ such that Eq.~(\ref{const}) is satisfied, Eq.~(\ref{Eq:IntermediateTDSE}) reduces to
\begin{equation}
	i\frac{d\ket{\psi}}{d\tilde{t}} = \tilde{H} \ket{\psi},
	\label{Eq:TransformedTDSE}
\end{equation}
with
\begin{equation}
	\tilde{H} = \frac{1}{2} \left( \begin{array}{cc} \tilde{\epsilon}(\tilde{t}) & \tilde{\Delta} \\ \tilde{\Delta} & - \tilde{\epsilon}(\tilde{t}) \end{array} \right),
	\label{Eq:RescaledLZHamiltonian}
\end{equation}
and
\begin{equation}
	\tilde{\epsilon}(\tilde{t}) = \frac{dt}{d\tilde{t}} \epsilon\left[t(\tilde{t})\right],
	\label{Eq:RescaledLZHamiltonian}
\end{equation}
where we treat $t$ as being a function of $\tilde{t}$. As a result, the effect of having a time-dependent gap $\Delta(t)$ is the same as the effect of having a fixed gap $\tilde{\Delta}$ but a modified sweep function $\tilde{\epsilon}(\tilde{t})$. Consequently the lowest-order effect of a temporal variation in $\Delta$ can be inferred from our results concerning the lowest-order effect of having a nonlinear sweep function $\epsilon(t)$.

\subsection{Illustrative examples}

As an example, we consider the following time dependence of the gap:
	\begin{equation}
		\Delta(t)=d_0 \left|\frac{t}{T}\right|^a, \qquad a>0.
	\end{equation}
	For $\tilde t>0$, the solution of Eq.~(\ref{IntegralForm}) gives the relation between $t$ and $\tilde t$ as follows
	\begin{equation}
		t =G(\tilde t)=
		\left[\frac{\tilde \Delta}{d_0}(a+1)T^a 
		\tilde t\right]^{1/(a+1)}.
	\end{equation}
Therefore, the function $\tilde\epsilon(\tilde t)$ in the resulting Hamiltonian~(\ref{Hnew}) is obtained explicitly.

As another example, let us take $\Delta(t) = \Delta_0 \left( 1 + \alpha \tanh t/T \right)$ and a linear sweep function $\epsilon(t)=vt$. We can set
\begin{equation}
	\frac{d\tilde{t}}{dt} = \left( 1 + \alpha \tanh\frac{t}{T} \right),
\end{equation}
or in other words
\begin{equation}
	\tilde{t} = t + T \alpha \ln \cosh \frac{t}{T},
	\label{Eq:ttilde}
\end{equation}
to get $\tilde{\Delta} = \Delta_0$. Note that we have set the integration constant in Eq.~(\ref{Eq:ttilde}) to zero, which sets the relation that $\tilde{t}=0$ when $t=0$. The modified function $\tilde{\epsilon}(\tilde{t})$ is then given by
\begin{equation}
	\tilde{\epsilon}(\tilde{t}) = \frac{vt(\tilde{t})}{1 + \alpha \tanh \frac{t(\tilde{t})}{T}}.
\end{equation}
The function $\tilde{t}(t)$ in Eq.~(\ref{Eq:ttilde}) cannot be inverted easily to obtain $t(\tilde{t})$ as a simple function. As a result, we cannot write $\tilde{\epsilon}(\tilde{t})$ explicitly.

\subsection{Perturbative formula}

Even if we cannot find an explicit expression for $\tilde{\epsilon}(\tilde{t})$, we can derive relations between the derivatives of $\tilde{\epsilon}(\tilde{t})$ with respect to $\tilde{t}$ and the derivatives of $\epsilon(t)$ and $\Delta(t)$ with respect to $t$, especially at the crossing point. The derivatives of $\tilde{\epsilon}$ with respect to $\tilde{t}$ allow us to calculate the adiabaticity and nonlinearity parameters in the transformed problem. As a result, the relations between the derivatives will allow us to infer the effect of a slowly time-dependent $\Delta(t)$ from our perturbative results for a weakly nonlinear $\epsilon(t)$.

For general time-dependent functions $\epsilon(t)$ and $\Delta(t)$, with the assumption that both $\Delta(t)$ and $d\epsilon/dt$ are always positive, we can set
\begin{equation}
	\frac{d\tilde{t}}{dt} = \frac{\Delta(t)}{\tilde{\Delta}},
\end{equation}
where $\tilde{\Delta}$ is a constant. We then obtain
\begin{equation}
	\tilde{\epsilon}(\tilde{t}) = \frac{\tilde{\Delta}\,\epsilon(t)}{\Delta(t)}.
\end{equation}
We can then evaluate the derivatives
\begin{widetext}
\begin{eqnarray}
	\frac{d\tilde{\epsilon}}{d\tilde{t}} \Bigg|_{\epsilon=0} & = & \left( \frac{d\tilde{\epsilon}}{dt} \times \frac{dt}{d\tilde{t}} \right) \Bigg|_{\epsilon=0}
	= \left( \left[ \frac{\tilde{\Delta}\,d\epsilon/dt}{\Delta(t)} - \frac{\tilde{\Delta}\,\epsilon(t) d\Delta/dt}{\Delta^2(t)} \right] \times \left[ \frac{\tilde{\Delta}}{\Delta(t)} \right] \right) \Bigg|_{\epsilon=0}
	= \tilde{\Delta}^2\left( \frac{d\epsilon/dt}{\Delta^2(t)} - \frac{\epsilon(t) d\Delta/dt}{\Delta^3(t)} \right) \Bigg|_{\epsilon=0}
	\nonumber \\
	& = & \frac{\tilde{\Delta}^2\, d\epsilon/dt}{\Delta^2(t)} \Bigg|_{\epsilon=0};
	\end{eqnarray}
	\begin{eqnarray}
	\frac{d^2\tilde{\epsilon}}{d\tilde{t}^2} \Bigg|_{\epsilon=0} & = & \left( \frac{d}{dt} \left( \frac{d\tilde{\epsilon}}{d\tilde{t}} \right) \times \frac{dt}{d\tilde{t}} \right) \Bigg|_{\epsilon=0}
	\nonumber \\
	& = & \tilde{\Delta}^3\Bigg( \frac{d^2\epsilon/dt^2}{\Delta^3(t)} - \frac{2(d\epsilon/dt) \times (d\Delta/dt)}{\Delta^4(t)}
	%
	%
	- \frac{(d\epsilon/dt) \times (d\Delta/dt)}{\Delta^4(t)}
	%
	%
	- \frac{\epsilon(t) d^2\Delta/dt^2}{\Delta^4(t)} + \frac{3 \epsilon(t) (d\Delta/dt)^2}{\Delta^5(t)} \Bigg) \Bigg|_{\epsilon=0}
	\nonumber \\
	& = & \tilde{\Delta}^3 \left( \frac{d^2\epsilon/dt^2}{\Delta^3(t)} - \frac{3(d\epsilon/dt) \times (d\Delta/dt)}{\Delta^4(t)} \right) \Bigg|_{\epsilon=0};
	\end{eqnarray}
	\begin{eqnarray}
	\frac{d^3\tilde{\epsilon}}{d\tilde{t}^3} \Bigg|_{\epsilon=0} & = & \left( \frac{d}{dt} \left( \frac{d^2\tilde{\epsilon}}{d\tilde{t}^2} \right) \times \frac{dt}{d\tilde{t}} \right) \Bigg|_{\epsilon=0}
	\nonumber \\
	& = & \tilde{\Delta}^4 \Bigg( \frac{d^3\epsilon/dt^3}{\Delta^4(t)} - \frac{3(d^2\epsilon/dt^2) \times (d\Delta/dt)}{\Delta^5(t)}
	%
	%
	- \frac{3(d^2\epsilon/dt^2) \times (d\Delta/dt)}{\Delta^5(t)}
	%
	%
	- \frac{3(d\epsilon/dt) \times (d^2\Delta/dt^2)}{\Delta^5(t)}
	\nonumber \\ & & \hspace{0.4cm}
	+ \frac{12(d\epsilon/dt) \times (d\Delta/dt)^2}{\Delta^6(t)}
	%
	%
	- \frac{(d\epsilon/dt) \times (d^2\Delta/dt^2)}{\Delta^5(t)}
	- \frac{\epsilon(t) d^3\Delta/dt^3}{\Delta^5(t)}
	\nonumber \\ & & \hspace{0.4cm}
	+ \frac{4\epsilon(t) (d^2\Delta/dt^2) (d\Delta/dt)}{\Delta^6(t)}
	%
	%
	+ \frac{3 (d\epsilon/dt) (d\Delta/dt)^2}{\Delta^6(t)}
	%
	%
	+ \frac{6 \epsilon(t) (d\Delta/dt) (d^2\Delta/dt^2)}{\Delta^6(t)}
	%
	%
	- \frac{18 \epsilon(t) (d\Delta/dt)^3}{\Delta^7(t)} \Bigg) \Bigg|_{\epsilon=0}
	\nonumber \\
	& = & \tilde{\Delta}^4 \Bigg( \frac{d^3\epsilon/dt^3}{\Delta^4(t)} - \frac{6(d^2\epsilon/dt^2) \times (d\Delta/dt)}{\Delta^5(t)}
	%
	%
	- \frac{4(d\epsilon/dt) \times (d^2\Delta/dt^2)}{\Delta^5(t)}
	%
	%
	+ \frac{15(d\epsilon/dt) \times (d\Delta/dt)^2}{\Delta^6(t)} \Bigg) \Bigg|_{\epsilon=0} .
\end{eqnarray}
\end{widetext}

These expressions can be used to evaluate the parameters $\chi_i$ ($i=2,3,...)$ and subsequently evaluate the transition probability. For example, if we take the case where $\epsilon(t)=vt$ and only the first derivative of $\Delta(t)$ is nonnegligible ($\Delta'=d\Delta/dt|_{\epsilon=0}$), we set $\tilde{\Delta}=\Delta(t)|_{\epsilon=0}=\Delta_0$ and obtain the rates
\begin{eqnarray}
	\frac{d\tilde{\epsilon}}{d\tilde{t}} \Bigg|_{\epsilon=0} & = & v,
	\nonumber \\
	\frac{d^2\tilde{\epsilon}}{d\tilde{t}^2} \Bigg|_{\epsilon=0} & = & - \frac{3v \Delta'}{\Delta_0},
	\nonumber \\
	\frac{d^3\tilde{\epsilon}}{d\tilde{t}^3} \Bigg|_{\epsilon=0} & = & \frac{15 v \times (\Delta')^2}{\Delta_0^2},
\end{eqnarray}
which in turn give the nonlinearity parameters
\begin{eqnarray}
	\chi_2 & = & - \frac{3 \Delta'}{v},
	\nonumber \\
	\chi_3 & = & \frac{15 (\Delta')^2}{v^2}.
	\label{Eq:VariableDeltaEquivalentNonlinearities}
\end{eqnarray}
Substituting these expressions for $\chi_2$ and $\chi_3$ in Eq.~(\ref{Eq:LZFormulaQuadraticCubic}), we obtain Eq.~(\ref{Eq:LZFormulaVariableDelta}).

\end{appendix}

\end{document}